\documentclass[preprint,journal]{vgtc}       

\ifpdf
  \pdfoutput=1\relax                   
  \usepackage{graphicx}                
  \DeclareGraphicsExtensions{.pdf,.png,.jpg,.jpeg} 
\else
  \usepackage{graphicx}                
  \DeclareGraphicsExtensions{.eps}     
\fi%

\graphicspath{{figures/}{pictures/}{images/}{./}} 

\usepackage{microtype}                 
\PassOptionsToPackage{warn}{textcomp}  
\usepackage{textcomp}                  
\usepackage{mathptmx}                  
\usepackage{times}                     
\usepackage{cite}                      
\usepackage{tabu}                      
\usepackage{booktabs}                  

\usepackage{comment}
\usepackage{subfigure} 
\usepackage{algpseudocode}
\usepackage[]{algorithm2e}
\usepackage[mathscr]{euscript}
\usepackage{amsfonts}
\usepackage{color, colortbl}
\usepackage{multirow}
\usepackage{setspace}
\usepackage{soul}
\usepackage[normalem]{ulem}

\newcommand{\Rspace}        {{\mathbb R}}

\newcommand{\Bscr}        {{\mathscr{B}}}

\newcommand {\mm}[1] {\ifmmode{#1}\else{\mbox{\(#1\)}}\fi}

\ieeedoi{10.1109/TVCG.2019.2934802}

\onlineid{1062}

\vgtccategory{InfoVis}
\vgtcpapertype{Algorithm/Technique}

\title{Persistent Homology Guided Force-Directed Graph Layouts}

\author{Ashley Suh, Mustafa Hajij, Bei Wang, Carlos Scheidegger, and Paul Rosen}
\authorfooter{
\item
 Ashley Suh is with the University of South Florida and Tufts University. E-mail: ashley.suh@tufts.edu.
\item
 Mustafa Hajij is with the Ohio State University. E-mail: hajij.1@osu.edu.
\item
 Bei Wang is with the University of Utah. E-mail: beiwang@sci.utah.edu. 
\item
 Carlos Scheidegger is with the University of Arizona. E-mail: cscheid@email.arizona.edu.
\item
 Paul Rosen is with the University of South Florida. E-mail: prosen@usf.edu.
}

\shortauthortitle{Suh \MakeLowercase{\textit{et al.}}: Force-Directed Graph Exploration Using Persistent Homology}

\abstract{Graphs are commonly used to encode relationships among entities, yet their abstractness makes them difficult to analyze. Node-link diagrams are popular for drawing graphs, and force-directed layouts provide a flexible method for node arrangements that use local relationships in an attempt to reveal the global shape of the graph. However, clutter and overlap of unrelated structures can lead to confusing graph visualizations. This paper leverages the persistent homology features of an undirected graph as derived information for interactive manipulation of force-directed layouts. We first discuss how to efficiently extract $0$-dimensional persistent homology features from both weighted and unweighted undirected graphs. We then introduce the interactive persistence barcode used to manipulate the force-directed graph layout. In particular, the user adds and removes contracting and repulsing forces generated by the persistent homology features, eventually selecting the set of persistent homology features that most improve the layout. Finally, we demonstrate the utility of our approach across a variety of synthetic and real datasets.} 


\teaser{
	\centering

    {\begin{minipage}[b]{.24\linewidth}
        \subfigure[F-R Force-Directed Layout\label{fig:teaser:before}]{{\includegraphics[width=\linewidth]{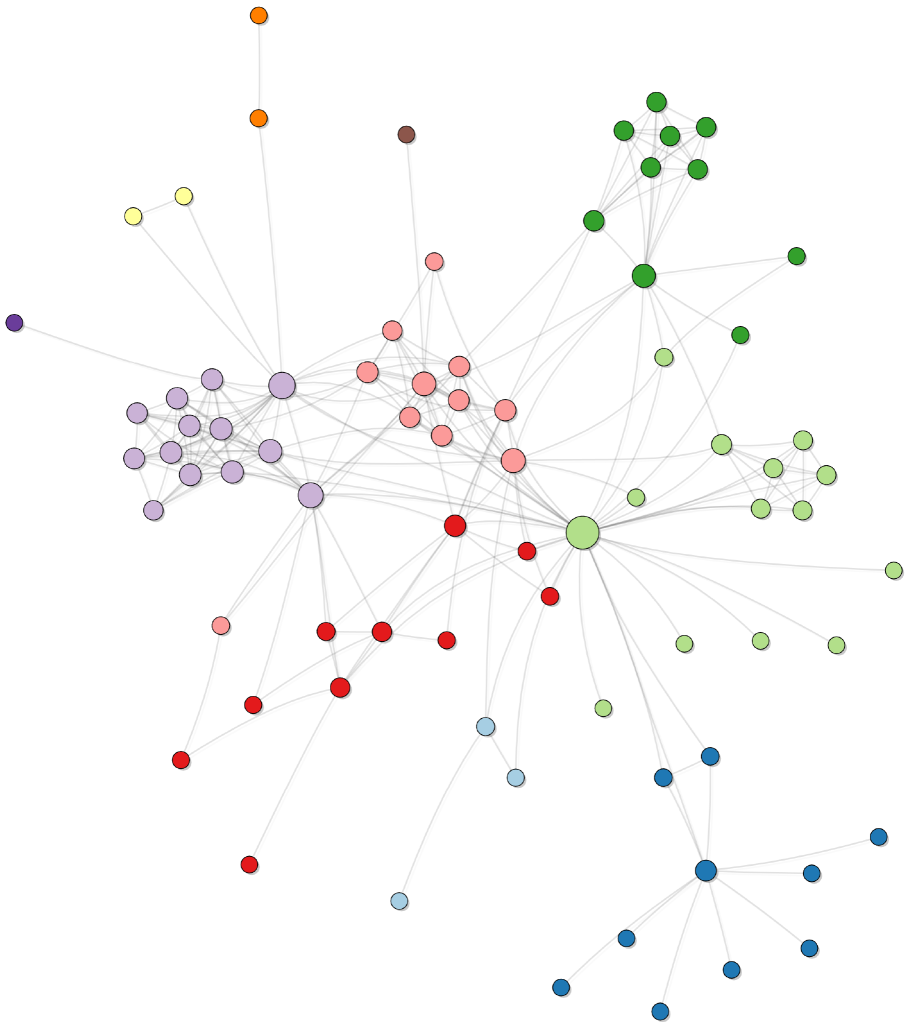}}}
        \vspace{10pt}
    \end{minipage}}
    \hfill
    {\begin{minipage}[b]{.18\linewidth}
        \centering
	    \subfigure[Contraction Only\label{fig:teaser:contraction}]{{\includegraphics[width=\linewidth]{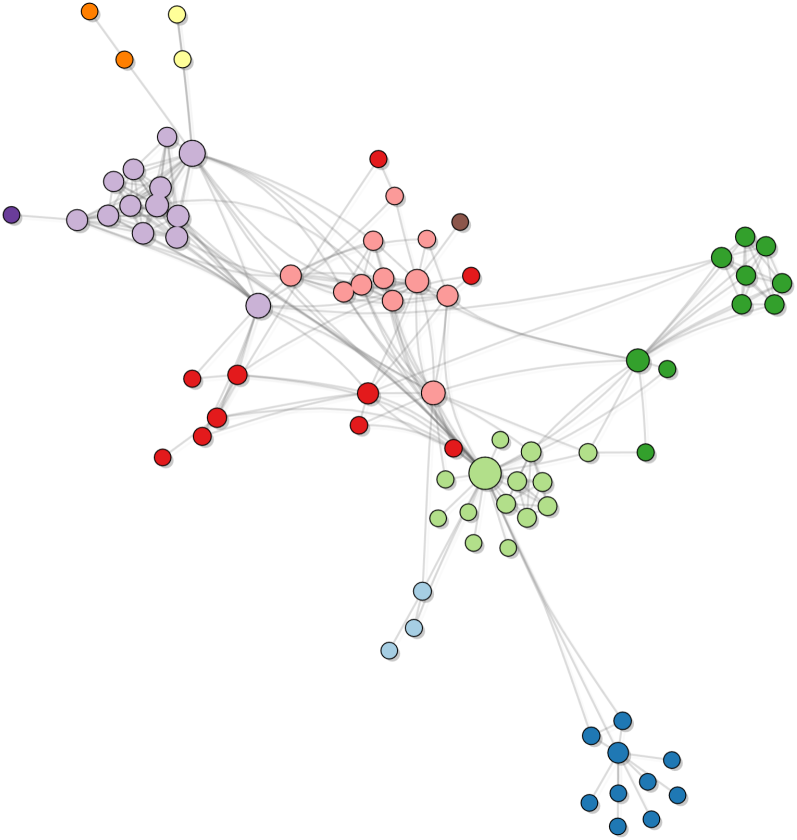}}}
	    
	    \vspace{-5pt}
	    \subfigure[Repulsion Only\label{fig:teaser:repulsion}]{{\includegraphics[width=\linewidth]{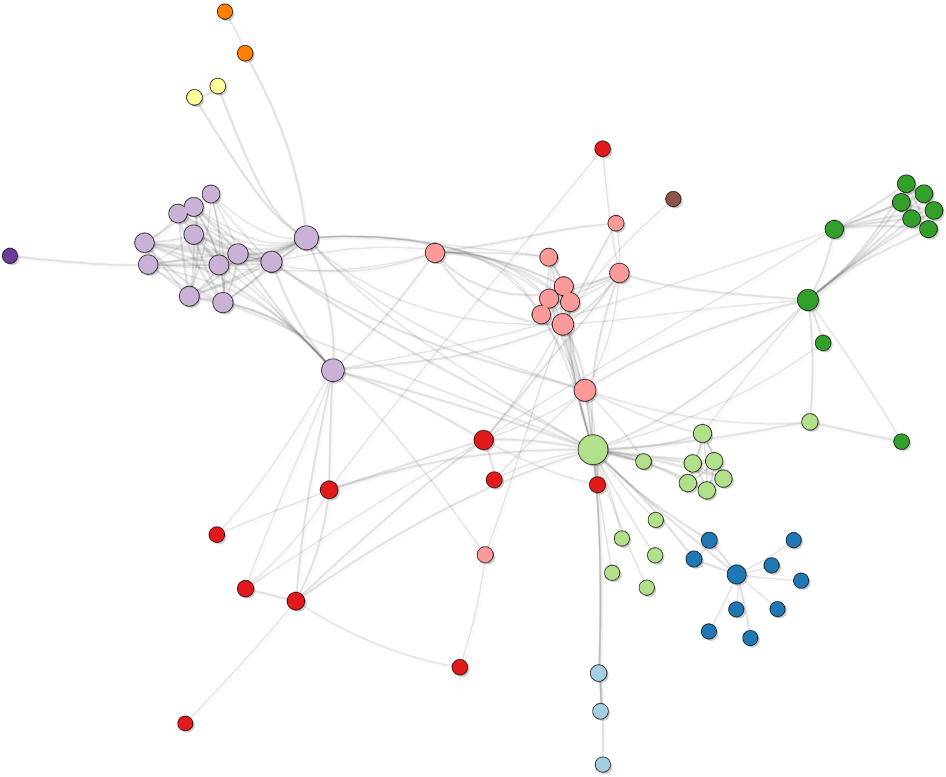}}}
    \end{minipage}}
    \hfill
    {\begin{minipage}[b]{.5\linewidth}
        \centering
	    \subfigure[Both Contraction and Repulsion\label{fig:teaser:both}]{{\includegraphics[height=2.65in]{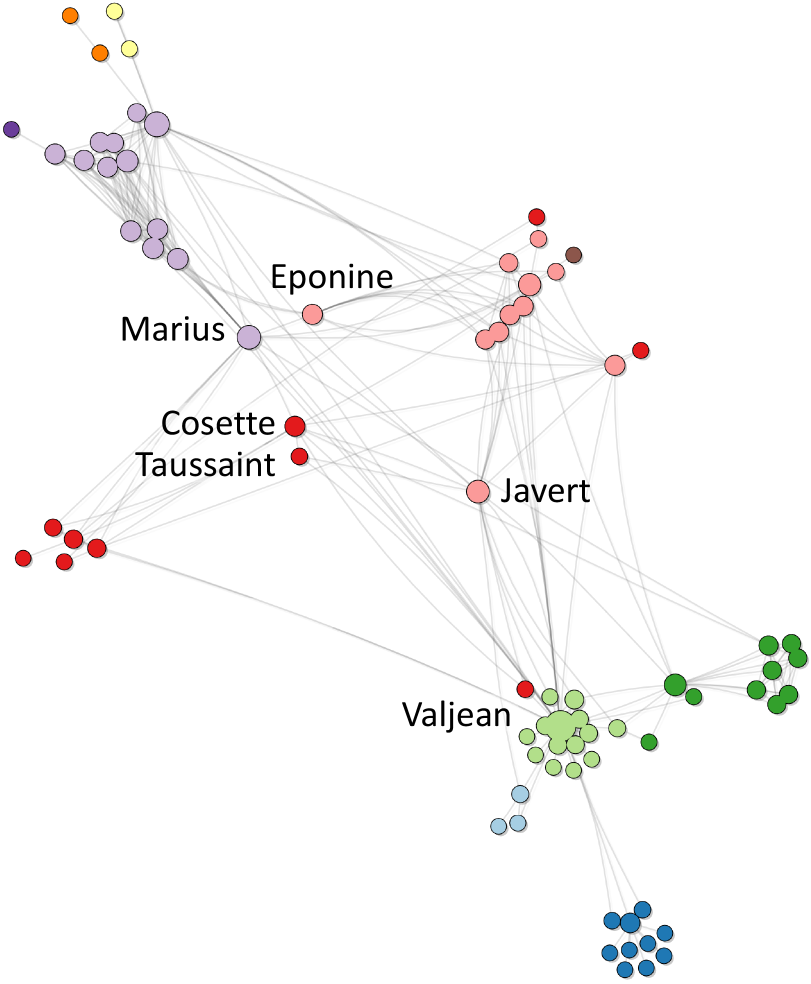}}}
	    \hspace{5pt}
	    \subfigure[Barcode\label{fig:teaser:barcode}]{{\includegraphics[height=2.65in]{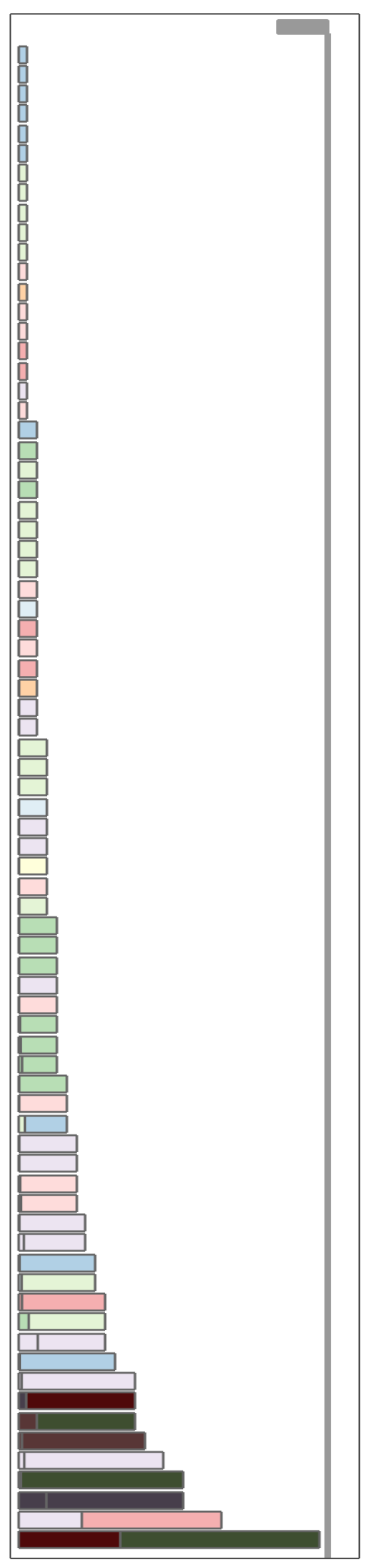}}}
    \end{minipage}}

    \caption{(a) The L\'es Miserables graph is drawn using a Fruchterman-Reingold (F-R) force-directed layout~\cite{fruchterman1991graph}. Our approach provides two mechanisms for interacting with the force-directed layout using (e) the persistence barcode. (b) The first mechanism contracts nodes of the graph associated with features of low significance or \emph{persistence}. (c) The second mechanism partitions the graph using user-selected features and repulses the nodes in different partitions from one another. (d) When combined, this approach allows interactively controlling the layout to emphasize user-selected aspects of the graph using persistent homology.}
    \label{fig:teaser}
}

\renewcommand{\paragraph}[1]{\vspace{3pt}\noindent\textbf{#1}}

\vgtcinsertpkg

\begin{document}

\maketitle

\setstretch{1}
\section{Introduction}

Graphs are ubiquitous for representing complex relationships between individuals or objects and are often used to model social interactions, energy grids, computer networks, brain connectivity, etc. The abstractness of graphs provides significant flexibility in visualization. However, dense, low-diameter subgraphs lead to confusing visualizations that appear as ``hairballs''. A good graph visualization should present \emph{structure} quickly and clearly, and support further investigation of the data. 

A key element of node-link diagrams is the layout algorithm that places nodes on the display (semi-)automatically. The problem of automatic graph layout has a rich literature, in which many approaches focus on finding an embedding of the graph by optimizing a readability metric~\cite{purchase2002metrics}, such as symmetry of the graph, lengths of the edges, or the number of edge crossings. A significant advancement was the realization that the use of derived information, such as node rank, graph distance, or approximate clustering, could improve graph layouts~\cite{GansnerKoutsofiosNorth1993, Noack2007, Dunne09improvinggraph}. However, many of these techniques either lack the ability to interactively manipulate the graph layout, or lack the temporal coherency of the layout necessary to make such interactions effective.

When considering graph layouts that support interactivity, perhaps the most popular (though not necessarily the \emph{best}) method is a force-directed or spring-mass layout~\cite{Gibson2012survey}, which converts the graph into a physical system of attractive springs and repulsive forces that iteratively minimize an energy function. These systems rely upon local relationships to reveal the overall shape in the graph. The result is a method that shows topological structures in certain graphs, particularly sparse ones. However, this approach often causes unrelated or distant topological structures to overlap or cross paths, making them difficult to differentiate. Some capacity to address this problem is provided through user interaction. Unfortunately, the interaction is most often through clicking and dragging individual nodes, which is ineffective for larger graphs and constrained by the forces applied to the graph layout. 

This paper addresses the interactive manipulation of force-directed graph layouts by leveraging \emph{persistent homology} (PH)~\cite {EdelsbrunnerHarer2008, Ghrist2008} as derived information for the visualization of undirected graphs. PH has recently been shown to be a robust descriptor of graphs~\cite{HajijWangScheidegger2018,rieck2017clique}, and it has a few key qualities that make it ideal for this application. First, the PH calculation extracts \textit{PH features}, in the form of $0$-dimensional homological groups, from a graph without the need to select parameters. Second, the PH features can be quantified and ranked according to their significance, known as \emph{persistence}. Third, they are invariant under small deformations, making them insensitive to noise and other small variations in data (e.g., removing a low-weight edge does not significantly change the graph)~\cite{HajijWangScheidegger2018}. Finally, the set of all PH features produces a compressed description of the graph that can be represented using a \emph{persistence barcode}~\cite{Ghrist2008}, which our approach uses as a graphical user interface to manipulate the graph via \emph{PH features}, instead of direct node manipulation.

In brief, our approach works as follows. We embed an undirected graph in a metric space by inducing a distance between all nodes. We extract the PH features (i.e., the $0$-dimensional homological features) of the metric space structure~\cite{EdelsbrunnerHarer2010} and sort them using their significance (i.e., persistence). Starting with a Fruchterman-Reingold force-directed layout~\cite{fruchterman1991graph}, we employ the PH features in two user-selectable ways. First, a selected PH feature can create a strong attractive force between the nodes that created the feature, causing them to contract (see \autoref{fig:teaser:contraction}). Second, a selected PH feature can be used to partition the graph into two subsets, which are repulsed from one another (see \autoref{fig:teaser:repulsion}). The user employs as many contractive or repulsive forces as desired, in order to emphasize graph elements of interest (see \autoref{fig:teaser:both}).

\paragraph{Contribution.} We demonstrate the usefulness of using $0$-dimensional PH features for controlling force-directed layouts. In summary: 1)~we discuss extracting PH features from both weighted and unweighted graphs; 2)~we introduce new forces into the layout that are derived from PH features; 3) we provide an interactive interface, based upon the persistence barcode, that allows users to interactively manipulate the layout using the PH features; and 4)~we evaluate the approach by comparing it to popular force-directed layouts and clustering algorithms.

\section{Prior Work}

\vspace{-5pt}
\paragraph{Graph Visualization.}
Graph visualization is a broad area, as demonstrated by von Landesberger et al.'s\ survey~\cite{von2011visual}. Our treatment focuses on approaches for drawing node-link diagrams~\cite{Gibson2012survey}, which are used to display graphs in popular visualization systems, including Gephi~\cite{bastian2009gephi}, NodeXL~\cite{HansenShneidermanSmith2010}, and Graphviz~\cite{ellson2002graphviz}. 

The first automated technique for laying out node-link diagrams was Tutte's barycentric coordinate embedding~\cite{Tutte1963}, followed by linear programming techniques~\cite{GansnerKoutsofiosNorth1993}, force-directed/mass-spring embeddings~\cite{fruchterman1991graph, Hu2005}, embeddings of the graph metric~\cite{GansnerKorenNorth2005}, and techniques exploiting linear-algebraic properties of the connectivity structures~\cite{BrandesPich2007, KhouryHuKrishnanScheidegger2012, Koren2003, KorenCarmenHarel2002}. Hybrid approaches, such as TopoLayout~\cite{archambault2007topolayout}, analyzed graph topology to identify the best type of graph embedding. Recent work using stress majorization has introduced the ability to add constraints that enable those layouts to highlight certain properties, such as stars, clusters, or circles~\cite{wang2018revisiting}. The more challenging problem of visualizing multivariate networks has been addressed through visual analytics approaches~\cite{van2014multivariate}.

Edge clutter presents a challenging problem for node-link diagrams. For denser graphs, edge bundling can reduce clutter by routing graph edges to the same portion of the screen~\cite{HoltenVanWijk2009}. In terms of quality, divided edge bundling~\cite{SelassieHellerHeer2011} produces high-quality results, whereas hierarchical edge bundling~\cite{GansnerHuNorthScheidegger2011} scales to millions of edges. There are also localized versions of edge bundling~\cite{wong2003edgelens} and filtering~\cite{tominski2006fisheye,hurter2011moleview} that adapt the display of edges based upon a user-selected region of interest. 

Other visual metaphors have been proposed to reduce overall clutter, ranging from relatively conservative proposals, such as replacing nodes with motifs~\cite{DunneShneiderman2013} based on graph topology or modules~\cite{DwyerRicheMarriotMears2013}, to more aggressive forms, such as variants of matrix diagrams~\cite{DinklaWestenbergWijk2012} and abstract displays of graph statistics~\cite{KairamMacLeanSavvaHeer2012}. 

When displaying a large dataset, it is natural to question the hard visual limits for graphs. Popular approaches such as pixel-based visualizations~\cite{Keim2000, KeimSchneidewindSips2007} encode large amounts of data within small rectangles or display pixels. Space-filling curves have also been used to build pixel-based graph visualizations~\cite{MuelderMa2008}. Furthermore, using visual boosting~\cite{OelkeJanetzkoSimon2011} tailored to network data may further reveal hidden information.

Research into interactive manipulation of force-directed node layouts includes interaction techniques~\cite{herman2000graph,von2011visual} such as panning and zooming, which are used to focus on regions of interest in a  graph. In addition to interacting directly with nodes in force-directed layouts~\cite{fruchterman1991graph, Hu2005}, approaches have included hierarchical layout constructions~\cite{henry1991interactive} and exploration~\cite{archambault2007grouse, archambault2008grouseflocks}, fisheye lenses~\cite{sarkar1992graphical,wang2019structure}, interactive refinement of automatic layouts~\cite{frohlich1994demonstration}, or constraint-based optimization~\cite{ryall1997interactive}.

\paragraph{Persistent Homology and Graphs.}
PH is an emerging tool in studying complex graphs~\cite{DonatoPetriScolamiero2012, ELuYao2012, horak2009persistent, PetriScolamieroDonato2013, PetriScolamieroDonato2013b}, including collaboration networks~\cite{BampasidouGentimis2014, carstens2013persistent} and brain networks~\cite{CassidyRaeSolo2015, DabaghianMemoliFrank2012, LeeChungKang2011b, LeeChungKang2011, LeeKangChung2012, LeeKangChung2012b, PirinoRiccomagnoMartinoia2015}. PH has recently been used in the visualization community for graph analysis targeting clique communities~\cite{rieck2017clique} and time-varying graphs~\cite{HajijWangScheidegger2018}. Many of these techniques take similar approaches, using PH to summarily analyze, quantify, and compare graphs. In contrast, our approach focuses on using PH to enable interactive manipulation of the graph layout.

\begin{figure}[!b]
    \centering
    
    \subfigure[Conversion from Graph to Metric Space Embedding\label{fig:filtration_example:metric}]{\includegraphics[width=0.75\linewidth]{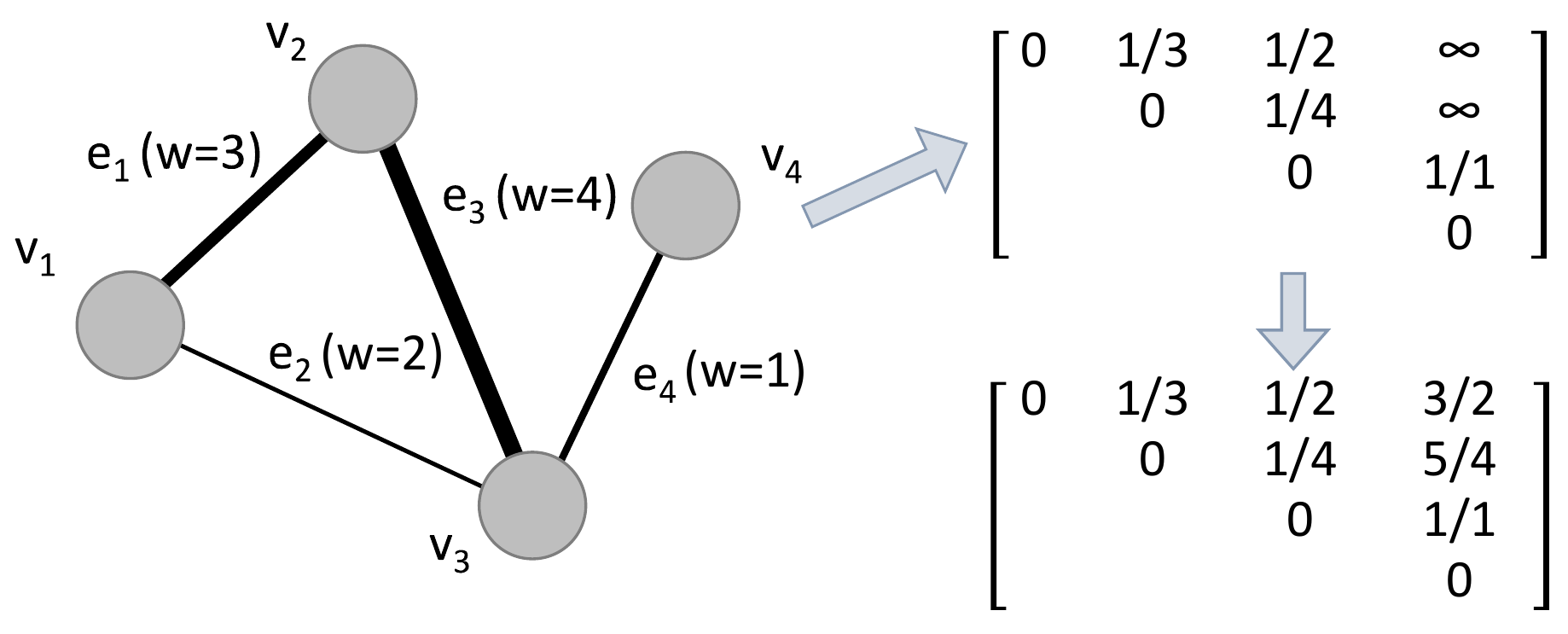}}
    \subfigure[Conceptual Construction of PH Features Using Growing Balls\label{fig:filtration_example:conceptual}]{\includegraphics[trim= 0 220pt 0 0, clip, width=0.925\linewidth]{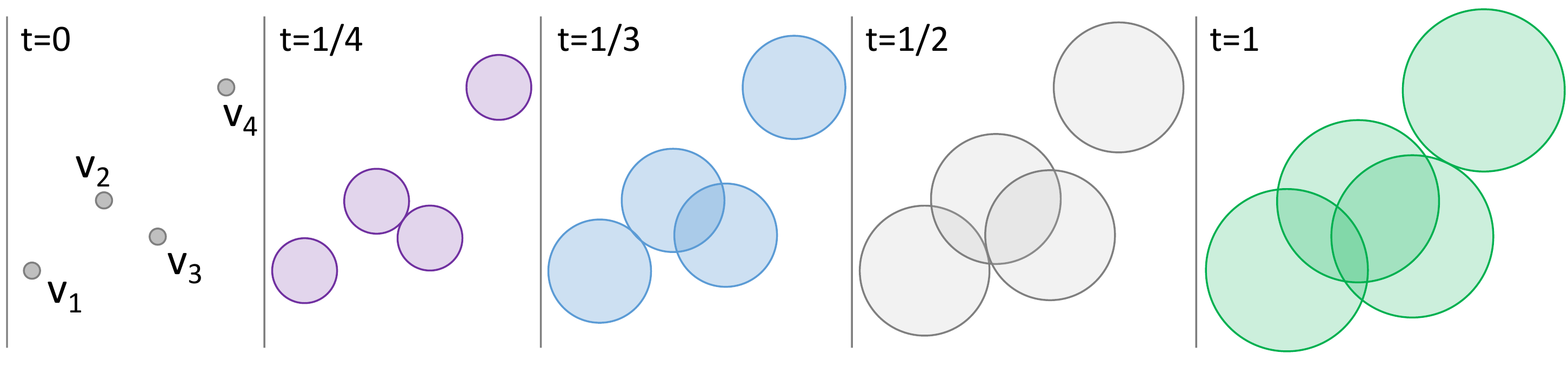}}
    \subfigure[Computational Construction of PH Features Using 1-Simplicies\label{fig:filtration_example:filtration}]{\includegraphics[trim= 0 0 0 215pt, clip, width=0.925\linewidth]{figures/filtration.pdf}}

    \caption{Example of extracting $0$-dimensional PH of a graph. (a) Given an undirected graph $G$ with edge weights $w$, we obtain a metric space representation by converting weights to distances by using $d=1/w$ and completing the metric space using shortest-path distance. (b) Conceptually, components, or PH features, are formed around each point in the metric space. Balls grow around the points in the metric space to identify the diameter $t$ at which components merge into larger components. (c) A filtration is constructed from $G$ by adding edges when two balls intersect. When two components merge into one, the bar associated with one of the component in the persistence barcode (bottom) terminates.}
    \label{fig:filtration_example}
\end{figure}

\section{Persistent Homology of a Graph}
\label{sec:extractPH}

We first provide a theoretic framing for our approach that is grounded in PH. In \autoref{sec:computePH} we will discuss how the restricted form of PH used in this paper is a special case of single-linkage hierarchical clustering.

To extract PH features from a graph, we apply PH to a metric space representation of the graph~\cite{HajijWangScheidegger2018}. See~\cite{EdelsbrunnerHarer2008} for an introductory survey and~\cite{EdelsbrunnerHarer2010} for a formal treatment of PH.

In algebraic topology, $0$-dimensional homology groups of a graph describe the connected components of a metric space at a \textit{single spatial resolution}. In this paper, we use a multiscale notion of homology, persistent homology (PH), to describe the evolution of features of the space at different spatial resolutions.

Given a weighted graph $G = (V, E, w)$ with positive edge weights defined by $w: E \to \Rspace$, our first step is to associate the graph $G$ with a metric space representation. 
Considering the \emph{inverse}\footnote{The inverse of the edge weight between two nodes $1/w(x,y)$ captures the dissimilarity between them.} of the positive edge weight as the length of an edge, a classical \emph{shortest path} metric $d$ is defined on $G$, where the distance between each pair of nodes $x, y \in G$ is the shortest path between them.  This metric can be computed using Dijkstra's algorithm~\cite{Dijkstra1959}. See \autoref{fig:filtration_example:metric} for an illustration. The remainder of the algorithm operates only on the metric space, no longer considering the original graph. 

Every node in $G$ corresponds to a point in the metric space. To compute the $0$-dimensional PH of $G$, we apply a simple geometric construction on its metric space representation. Consider the set of balls centered at every point in the metric space with a diameter $t$. We keep track of how the components of the union of balls evolve as $t$ increases from $0\rightarrow\infty$. As $t$ increases, the unions of balls form components in a hierarchical fashion.   

Considering \autoref{fig:filtration_example:conceptual}, starting with each point as a component when $t=0$, as $t$ increases, the number of components decreases by one when two components merge---formally, this is referred to as a \emph{topological event}\footnote{In this context, topology refers to the homology groups of the metric space, not to be confused with graph topology.}. At $t=1/4$, the balls representing $v_2$ and $v_3$ touch, causing the merging of the components $\{v_2,v_3\}$. At $t=1/3$, the balls representing $v_1$ and $v_2$ touch, causing the merging of sets $\{v_1\}$ and $\{v_2,v_3\}$. At $1/2$, $v_1$ and $v_3$ touch. However, they are already part of the same component $\{v_1,v_2,v_3\}$, meaning no merging occurs. Finally, at $t=1$, $v_3$ and $v_4$ touch, merging the components $\{v_1,v_2,v_3\}$ and $\{v_4\}$.

\textit{A PH feature corresponds to the birth (appearance) and death (merging) of a component (union of balls) in the metric space.} The \emph{birth} of a component is the diameter when the component appears. In our case, all points appear simultaneously with diameter $0$. The \emph{death} is the diameter at which a component disappears; that is, when two components merge, one will disappear by joining the other (the choice of which disappears is discussed in the next section). The lifetime of a component (i.e., its death time minus its birth time) is its \emph{persistence}. 

The PH features associated with $G$ are placed into a \emph{persistence barcode}~\cite{Ghrist2008}, which consists of a collection of bars, each corresponding to a single PH feature, whose starting and ending points correspond to the birth time and the death time of its associated component, with a width proportional to its persistence. 
See \autoref{fig:filtration_example:filtration} for an illustration.

\section{Computation of Persistent Homology}

The restricted form of PH used in this paper is functionally equivalent to single-linkage clustering. Therefore, the PH of the graph can be calculated by finding the minimum spanning tree (MST) of the graph using Kruskal's algorithm~\cite{kruskal1956shortest}.

\subsection{Fast Computation of the 0-Dimensional Barcode}
\label{sec:computePH}

In using Kruskal's algorithm to compute the $0$-dimensional barcode of the graph $G$, we suppose for simplicity that $G$ is connected. However, if it is not, each connected component of the graph is processed independently. In a nutshell, our algorithm consists of computing the MST $T$ of $G = (V, E, w)$ based on its metric space embedding using edge lengths $1/w$.

Let $V=\{v_1,\cdots,v_m\}$ be the node set of $G$ and let $E=\{e_1,\cdots,e_n\}$ be its edge set sorted in increasing order with respect to $1/w$.

The algorithm starts by creating an empty spanning tree. It then creates components $C_i$ and bars $\Bscr_i$, one per graph node. Each bar in the persistence barcode is represented by a pair of real numbers $(birth,death)$, initially $birth=0$ for each node as its own component. The second step of the algorithm looks at the edges one at a time, ordered by increasing $1/w_i$. For each $e_i$, we check if nodes of this edge belong to two different components. If this is the case, then we merge the components and set the $death$ of one to be the $1/w_i$. The choice of which component dies is arbitrary and does not influence our result. The \emph{persistence} of the component that dies is its death time minus its birth time. The appearance and the disappearance of such a component gives rise to a bar $(0, 1/w_i)$ in the persistence barcode. This step can be performed efficiently using the \emph{disjoint set} data structure. 

\vspace{-8pt}
\LinesNumbered 
\begin{algorithm}
 \KwData{A weighted graph $G = (V, E, w)$}
 \KwResult{Minimum spanning tree $T$ and $0$-dim barcode $\Bscr$}
 Create an empty spanning tree $T=\{\}$ \\
 \ForEach{node $v_i$}{ 
 	Create a component $C_i$ = $\{v_i\}$ \\
 	Create a bar $\Bscr_i$ with $birth=0$ and $death=\infty$
  }
 \ForEach{edge $e_i=(u,v)$ in $E$}{
 	
 	\If{$C_u$ and $C_v$ are different components}{
    	Merge $C_u$ and $C_v$ \\
        Set the $death$ of $\mathscr{B}_u$ to $1/w(e_i)$\\
        Add $e_i$ to the spanning tree $T$
    }
 }
\vspace{-8pt}
\end{algorithm}

The addition of an edge to the spanning tree coincides with the event of two components merging. Therefore, there is a one-to-one correspondence between edges of the MST $T$ and the $0$-dimensional barcode of $G$ with finite persistence.

\begin{figure}[!b]
	\centering

    \includegraphics[width=0.875\linewidth]{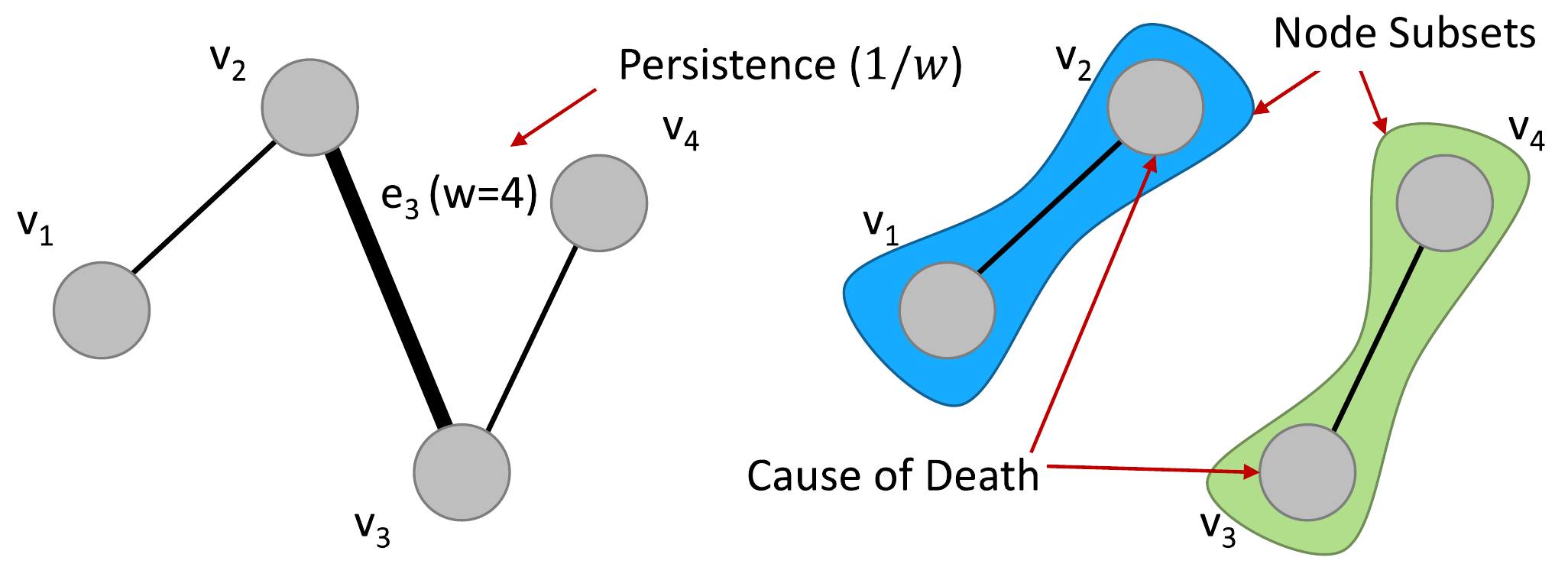} 

	\caption{Example of information extracted from a spanning tree (left) for edge $e_3$ from \autoref{fig:filtration_example:filtration} (the purple bar). The right shows the clusters created when a selected edge is removed from the spanning tree.}
	\label{fig.barcode_data}
\end{figure}

\subsection{Node Relationships with the Spanning Tree}
\label{sec:ph:st}

We relate information encoded by the MST to the graph $G$, which will later define modifications to the graph layout. Denote the MST as $T(V, E)$, where $E$ denote the edges in the tree. Deleting an edge $e=(u,v)$ from $E$ splits the tree $T$ into two sets, $V_u$ and $V_v$.

\begin{figure}[!ht]
    \centering

    \hspace{8pt}
    \begin{minipage}[c]{0.44\linewidth}
        \includegraphics[width=1\linewidth]{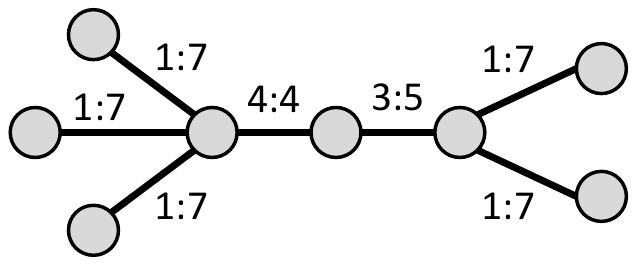}
    \end{minipage}
    \hfill
    \begin{minipage}[c]{0.44\linewidth}
        \vspace{-10pt}
        \caption{Example MST with associated subset ratios. Nodes toward the periphery have low ratios, whereas more central nodes have balanced ratios.}
        \label{fig:split_ratio}
    \end{minipage}
    \hspace{8pt}
\end{figure}

\textit{Each} PH feature (i.e., a bar in the barcode) is associated with the following information, as illustrated in \autoref{fig.barcode_data}:
\vspace{-5pt}
\begin{itemize}
  \setlength{\itemsep}{2pt}
  \setlength{\parskip}{0pt}
  \setlength{\parsep}{0pt}
    \item For our purpose, we visualize each bar in the persistence barcode as an interval $(0, w)$ instead of $(0, 1/w)$. Such a visualization emphasizes high weight edges as long bars\footnote{This visualization is different from the conventional PH approach; however it is  justified as edges with higher weights $w$ are considered more important in our setting.}. Under an abuse of notation, the \emph{persistence measure} of such a bar is assigned $w$.
    \item The \emph{cause of death}, $u$ and $v$, are the nodes of the edge that cause the components to merge. These nodes will be used to modify the graph layout to the reflect PH feature selection.
    \item The \emph{subsets of nodes}, $V_u$ and $V_v$, represent the sets of connected nodes after the removal of the edge from the MST. These sets will also be important when updating the graph layout.
    \item The \emph{subset ratio} is a measure of the number of nodes, $|V_u|:|V_v|$ in the two subsets of nodes.  It is a measure of centrality within the MST. For example, in \autoref{fig:split_ratio}, an example MST is augmented with the subset ratios. Edges in the fan-like areas to the left and right have low ratios, $1$:$7$. The 2 central edges have a more balanced ratio, $4$:$4$ and $3$:$5$, indicating that they are more central to the MST. The distribution of subset ratios is entirely data dependent. However, we observed in our examples that low ratios are far more common than balanced ratios.
\end{itemize}

\subsection{Application to Unweighted Graphs}
\label{sec:phUnweighted}

The calculation described above requires a weight on the edges of the graph. For unweighted graphs, a similarity measure, such as the Jaccard index~\cite{jaccard1901etude}, can take the place of edge weight. 

Our procedure first gathers for each node a neighborhood or ego graph~\cite{newman2003ego} within a user-selectable number of hops. For example, a 1-hop ego graph will contain adjacent neighbors, whereas a 2-hop ego graph will contain both 1-hop neighbors and nodes adjacent to the 1-hop neighbors. The ideal number of hops is data dependent. We used 1-hop for denser graphs and 2- or 3-hops for sparser graphs. Then, given an edge $e=(v_i,v_j)$, with neighborhood graphs $N_i$ and $N_j$, the Jaccard index between those nodes is $J(v_i,v_j) = {{|N_i \cap N_j|}\over{|N_i \cup N_j|}}$. The edge weight becomes $w(e)=J(v_i,v_j)$. In this way, the Jaccard index provides the similarity between the neighbors of 2 connected nodes. Finally, we proceed with the approach described in \autoref{sec:extractPH}.  

Other measures, such as edge centrality~\cite{girvan2002community}, can be used in place of the Jaccard index, to highlight different aspects of the graph. The only requirement is that the measure must provide a weight on the edges.

\section{Visual Design and Interaction Design}
\label{sec:vis-design}

Our design goal is to use the PH of a graph as a control to manipulate its force-directed layout. We use a simple interactive interface to provide contextual information and enable fast manipulation of the layout. 

\begin{figure}[!b]
    \centering

    \subfigure[Force-directed layout\label{fig:graph-drawing:a}]{\includegraphics[width=0.4\linewidth]{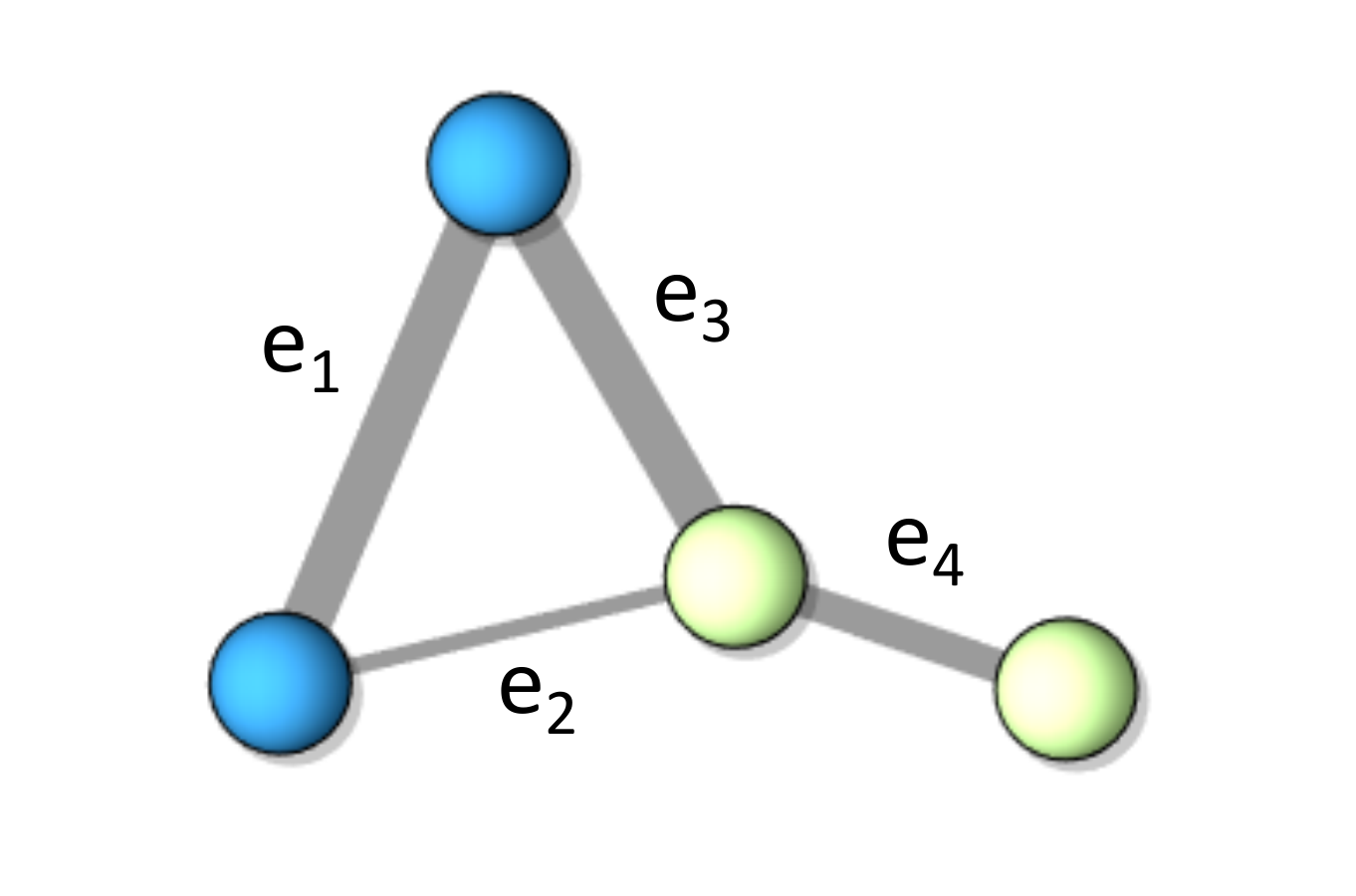}}
    \subfigure[Interactive barcode\label{fig:graph-drawing:b}]{\includegraphics[width=0.475\linewidth]{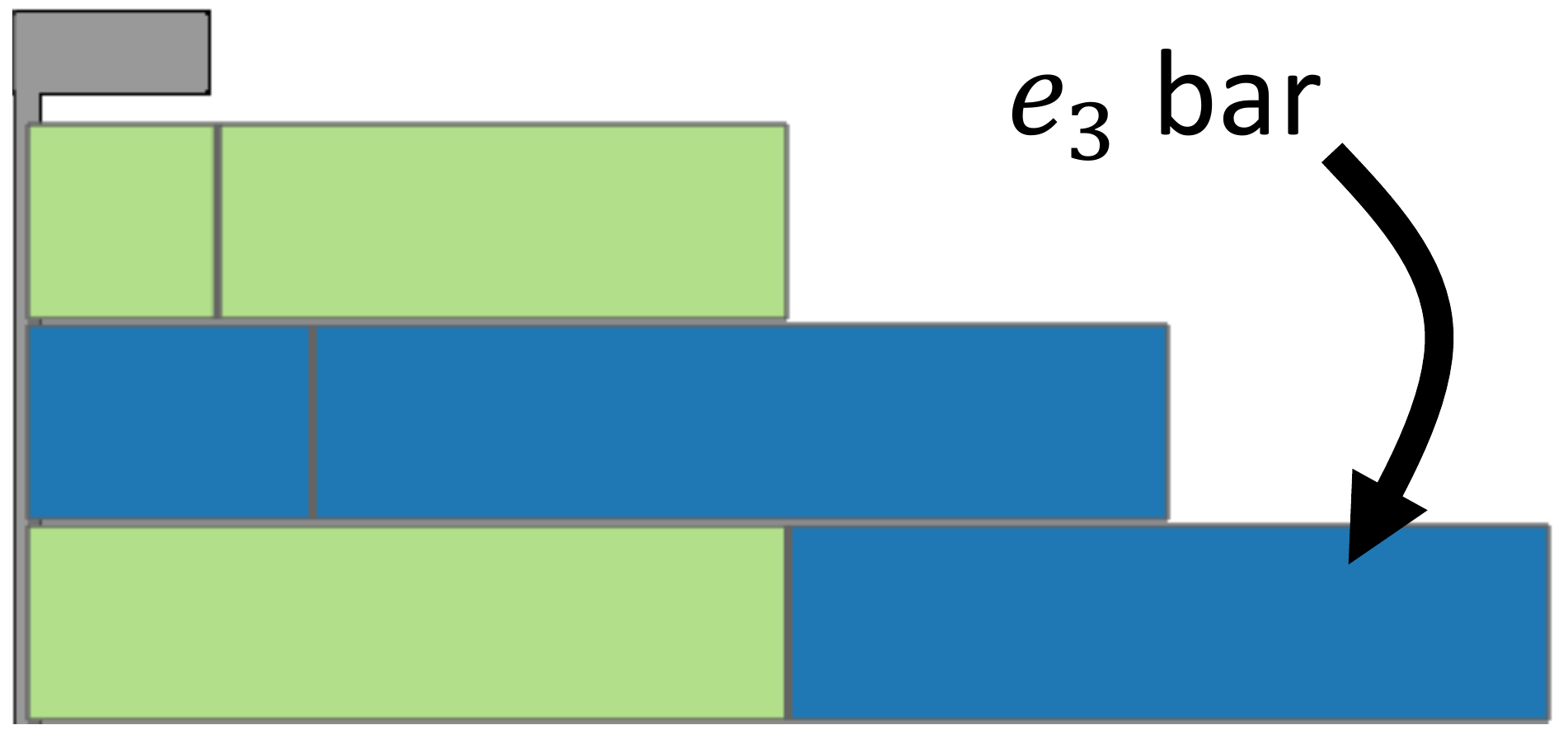}}

    \subfigure[Repulsive\label{fig:graph-drawing:c}]{\includegraphics[width=0.27\linewidth]{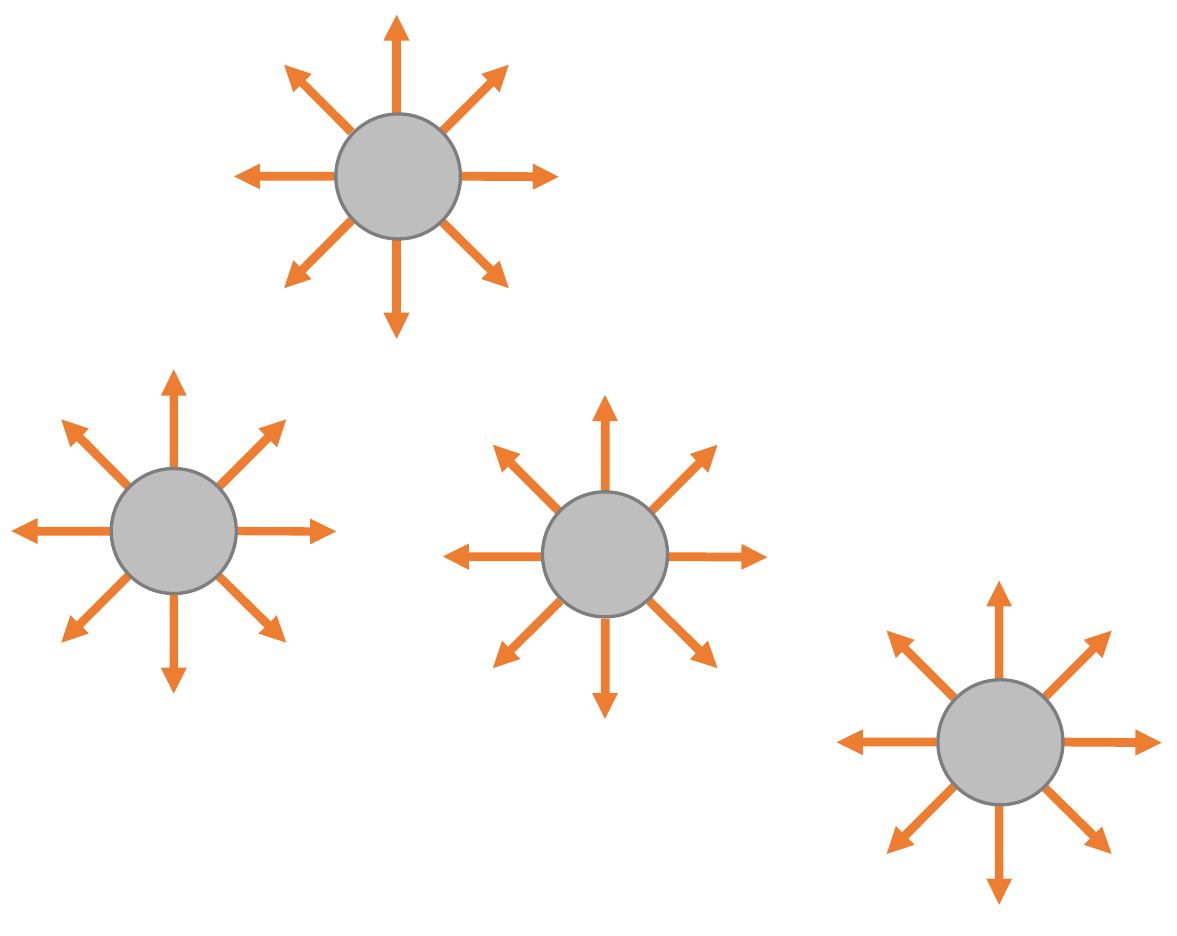}}
    \hspace{5pt}
    \subfigure[Spring attractive\label{fig:graph-drawing:d}]{\includegraphics[width=0.27\linewidth]{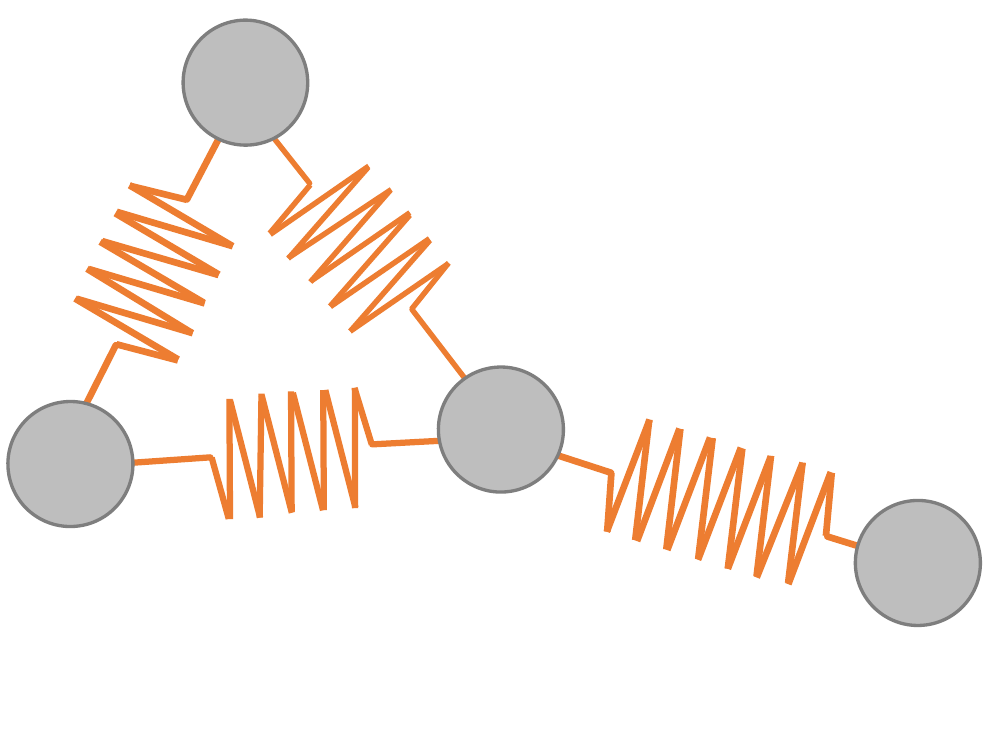}}
    \hspace{5pt}
    \subfigure[Centering\label{fig:graph-drawing:e}]{\includegraphics[width=0.27\linewidth]{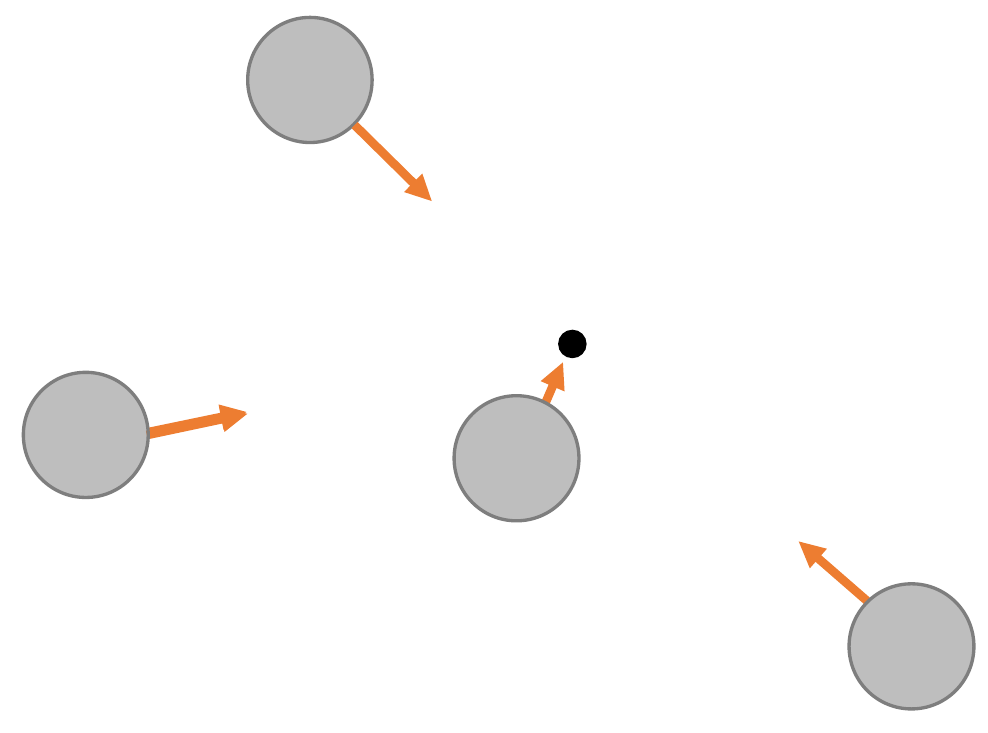}}

    \caption{The force-directed layout (a) for the graph in \autoref{fig:filtration_example} is constructed using (c) repulsive forces, (d) spring attractive forces, and (e) a centering force. The interactive barcode (b) is used to manipulate the display.}
    \label{fig:graph-drawing}
\end{figure}

\subsection{Graph Drawing}
\label{sec.vis.gd}

Our design uses a force-directed layout, with optional edge bundling. 

\paragraph{Force-Directed Layout.}
A graph is initially drawn with a Fruchterman-Reingold (F-R) force-directed layout~\cite{fruchterman1991graph}. 
The layout starts with three types of forces. The first force enables all nodes to repulse one another (\autoref{fig:graph-drawing:c}). The second force is a spring attraction for nodes connected by an edge (\autoref{fig:graph-drawing:d}). Finally, a weak attracting force draws all nodes toward the middle of the display, essentially centering the layout (\autoref{fig:graph-drawing:e}). The parameters for these forces, such as mass, force strength, and spring resting length, require manual tuning.

\paragraph{Edge Bundling.}
For analysis tasks that benefit from further clutter reduction~\cite{bach2017towards,mcgee2012empirical}, edge bundling can be optionally used to reduce the distractive impact of overlapping edges.
Since the layout is actively manipulated, we require an edge bundling implementation that is temporally coherent.  To accomplish this, we implemented force-directed edge bundling~\cite{HoltenVanWijk2009}. This technique subdivides edges and uses a variant of a force-directed layout to attract edges with similar proximity and direction. \autoref{fig:teaser} shows an example with edge bundling enabled.

\paragraph{Node Coloring.}
Some of our experimental graphs have categorical data attached to the nodes and are colored using a categorical color map. Other graphs contain nodes colored using node degrees.

\subsection{Persistence Barcode}
As mentioned in \autoref{sec:ph:st}, we visualize the persistence barcode in a unconventional way. A PH feature that is born at $0$ and dies at $1/w$ is visualized by a bar $(0, w)$ whose length indicates its persistence measure $w$. That is, the larger the bar, the more ``important'' the PH feature it represents.
We augment the barcode with additional visual encodings (\autoref{fig:graph-drawing:b}) to further guide interaction.

\paragraph{Subset Ratio.} 
Each bar is augmented with a vertical line, splitting it into two based upon the \emph{Subset Ratio}. For example, in \autoref{fig:graph-drawing:b} the bottom bar, representing $e_3$ from \autoref{fig:filtration_example} and~\ref{fig.barcode_data}, has a 50/50 split, because two nodes exist on either side of the edge in the MST (see \autoref{fig.barcode_data}).

\paragraph{Color.} 
For graphs with categorical data, each side of the split is colored based on the category of the \emph{Cause of Death} nodes for the PH feature.

\paragraph{Bar Sorting.}
Bars are sorted based upon two criteria. The first is \emph{persistence} from low to high, which helps to differentiate high persistence features from those with low persistence. If the persistence values of two bars are equal, then they are sorted by their \emph{Subset Ratios}, with bars having 50/50 ratios appearing lower in the barcode, since those are more central to the MST.

\begin{figure}[!t]
    \centering
    \subfigure[Contraction\label{fig:smallGraph:a}]{
        \begin{minipage}{0.45\linewidth}
            \centering
            \includegraphics[width=0.7\linewidth]{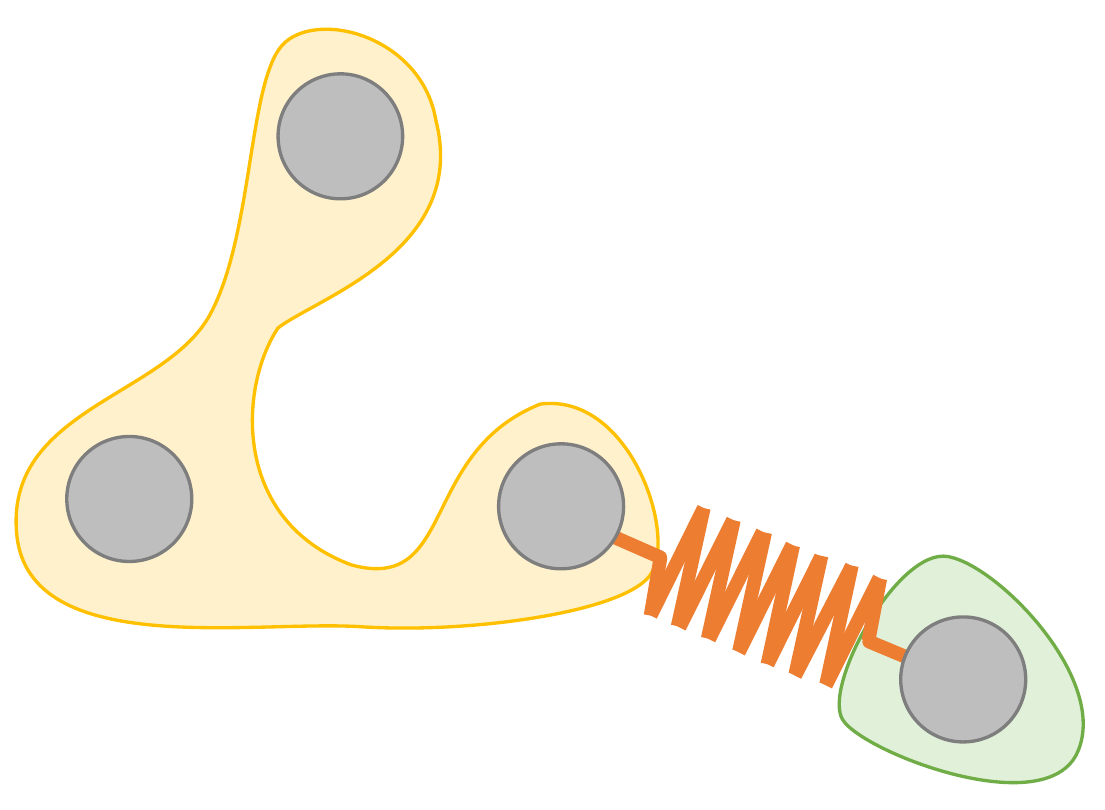}
        \end{minipage}
    }
    \subfigure[Repulsion\label{fig:smallGraph:b}]{
        \begin{minipage}{0.45\linewidth}
            \centering
            \includegraphics[width=0.7\linewidth]{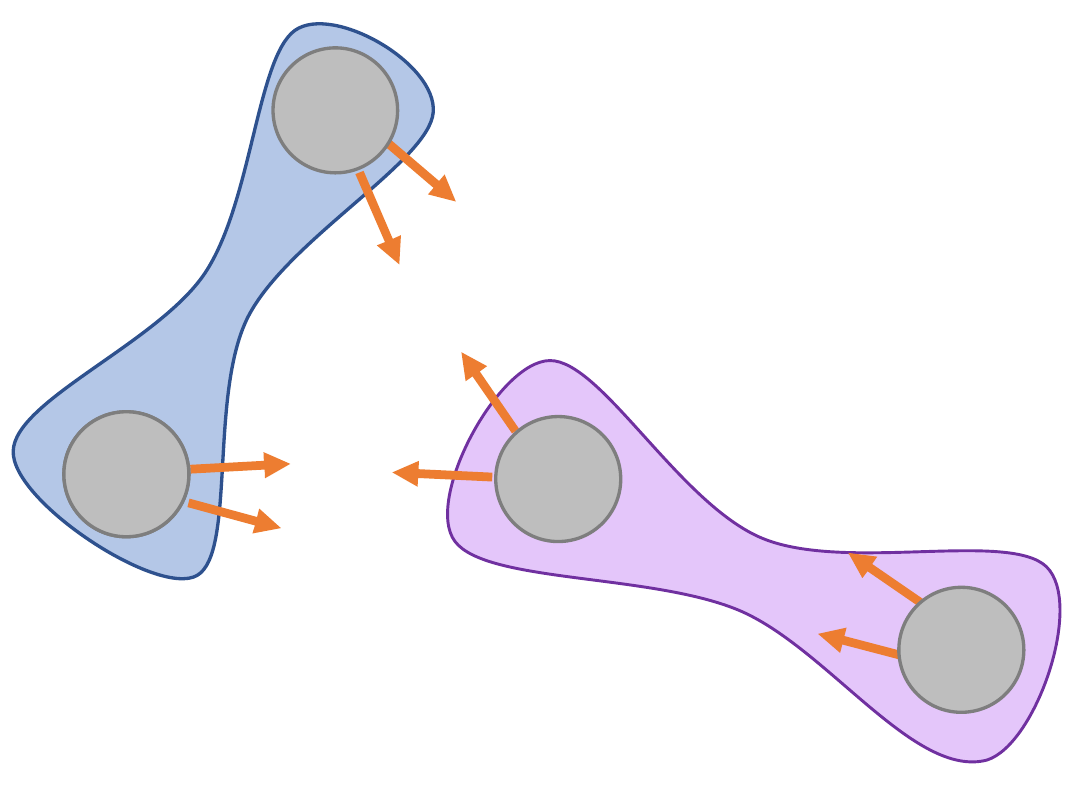}
        \end{minipage}
    }
    
    \subfigure[Contraction selection\label{fig:smallGraph:c}]{
        \begin{minipage}{0.45\linewidth}
            \centering
            \includegraphics[width=0.8\linewidth]{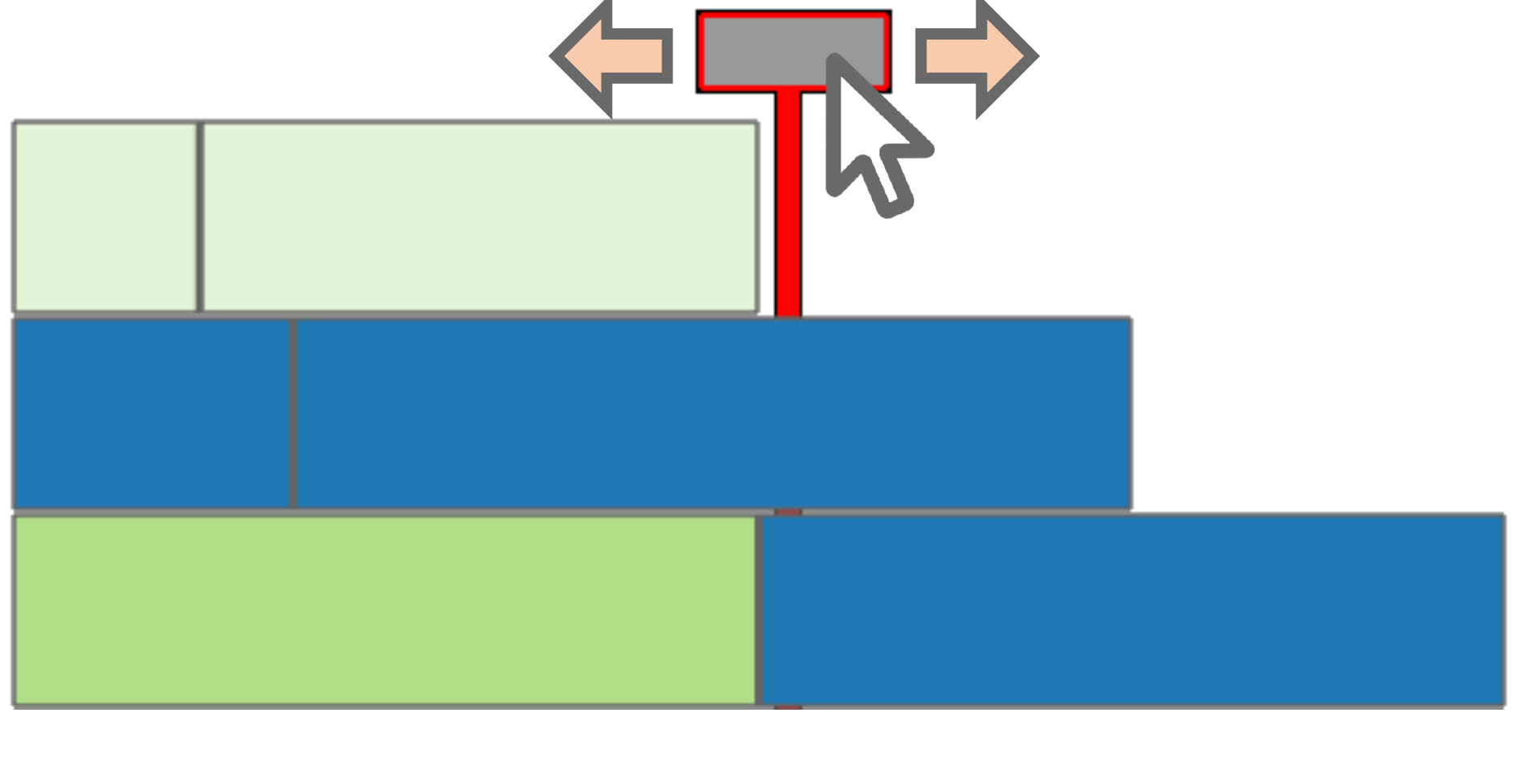}
        \end{minipage}
    }
    \subfigure[Repulsion selection\label{fig:smallGraph:d}]{
        \begin{minipage}{0.45\linewidth}
            \centering
            \includegraphics[width=0.8\linewidth]{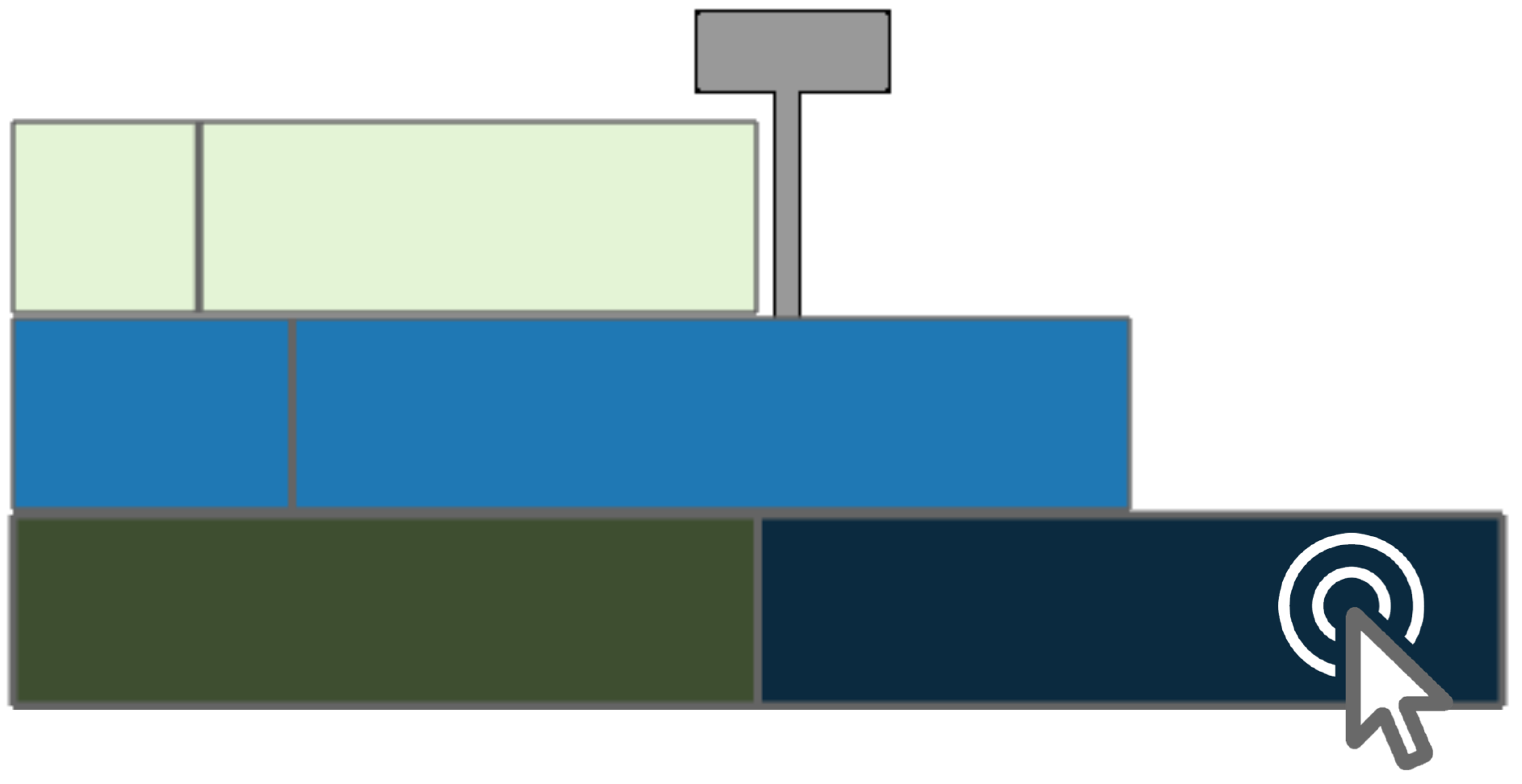}
        \end{minipage}
    }

    \subfigure[Adding contraction\label{fig:smallGraph:e}]{
        \begin{minipage}{0.45\linewidth}
            \centering
            \includegraphics[width=0.75\linewidth]{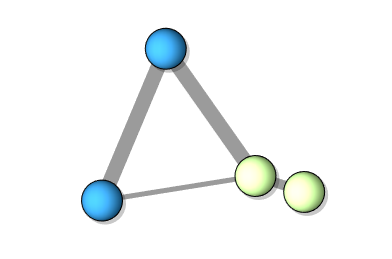}
        \end{minipage}
    }
    \subfigure[Adding repulsion\label{fig:smallGraph:f}]{
        \begin{minipage}{0.45\linewidth}
            \centering
            \includegraphics[width=0.75\linewidth]{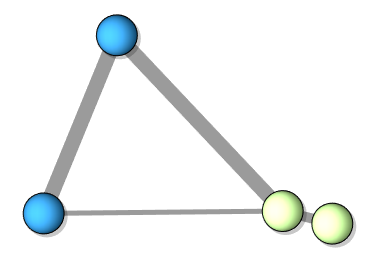}
        \end{minipage}
    }

    \caption{Illustration of forces applied to the graph in \autoref{fig:graph-drawing:a}. (a, c, and e): A contracting force applied to the layout. (b, d, and f): A repulsive force is then added to the layout from (e).}
    \label{fig:smallGraph}
\end{figure}

\subsection{Interaction Using the Persistence Barcode}

\paragraph{PH Feature Contraction.} 
In certain scenarios, it is desirable to shrink space allocated to nodes in the layout, making more room for other parts of the graph to be displayed. We provide a \textit{persistence simplification} tool that enables contraction of all PH features whose persistence is below a user-selected threshold. This interaction is done by dragging a filter bar at the top of the barcode; see the filter bar in red in \autoref{fig:smallGraph:c}. 
As the threshold is dragged left to right, PH features with persistence below that threshold will have their color washed out and their graph nodes contracted. A contraction is accomplished by adding a strong spring force between the \emph{cause of death} nodes for the event. The scale of this force is user selectable. \autoref{fig:smallGraph:a} illustrates this force and \autoref{fig:smallGraph:e} shows an example of the contraction.

\paragraph{PH Feature Repulsion.} In other situations, stretching out the graph can clear space for nodes and edges that might otherwise be overlapping and difficult to track. When a bar is individually selected, the bar color is darkened, and a strong repulsive force is added between the \emph{subsets of nodes} associated with that PH feature. The scale of the force is user selectable. The repulsion of the nodes from these two groups allows the layout to cluster.
For example, \autoref{fig:smallGraph:b} illustrates the force when the bottom bar in \autoref{fig:smallGraph:d} is selected for repulsion, causing the subsets to push apart. In other words, in \autoref{fig:smallGraph:b}, all the points in the blue region have an additional repulsion from all of the points in the purple region and vice versa. \autoref{fig:smallGraph:f} shows the result of adding this force to the graph.

\paragraph{Selecting Multiple Bars.} Multiple bars may be selected for contraction and repulsion, since the forces do not directly depend on each other. Whenever a new bar is selected, the force is simply added to the layout.

\begin{figure}[!b]
    \centering

    \hspace{2pt}
    \subfigure[Partition mixed in graph before (top) and after (bottom) repulsion\label{fig:contract_ex:a}]{
        \begin{minipage}{0.45\linewidth}
            \includegraphics[trim=150pt 5pt 50pt 40pt, clip, width=1\linewidth]{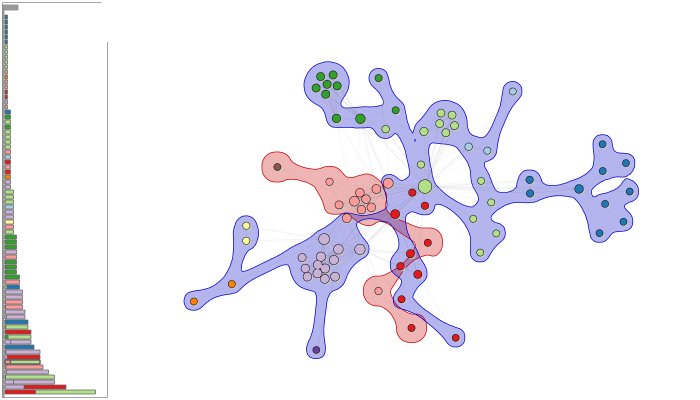}
            \includegraphics[trim=150pt 5pt 50pt 40pt, clip, width=1\linewidth]{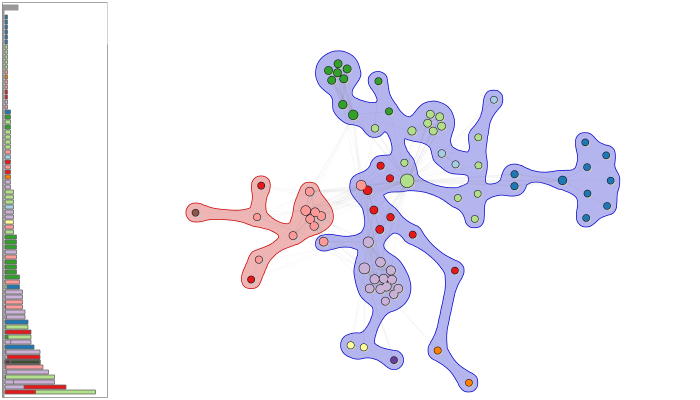}
        \end{minipage}
    } 
    \hfill
    \subfigure[Partition to be separated before (top) and after (bottom) repulsion\label{fig:contract_ex:b}]{
        \begin{minipage}{0.45\linewidth}
            \includegraphics[trim=150pt 5pt 50pt 40pt, clip, width=1\linewidth]{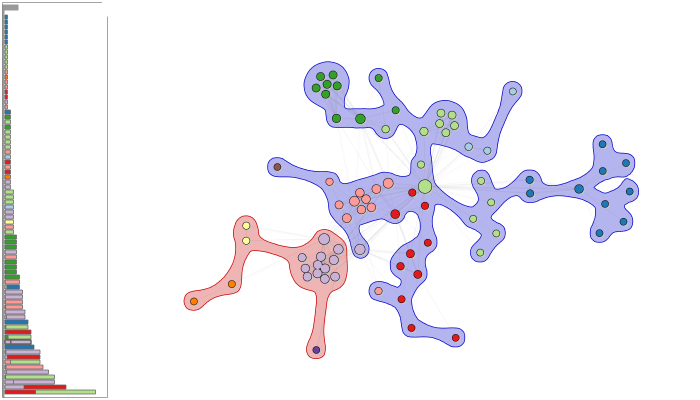}
            \includegraphics[trim=140pt 5pt 60pt 40pt, clip, width=1\linewidth]{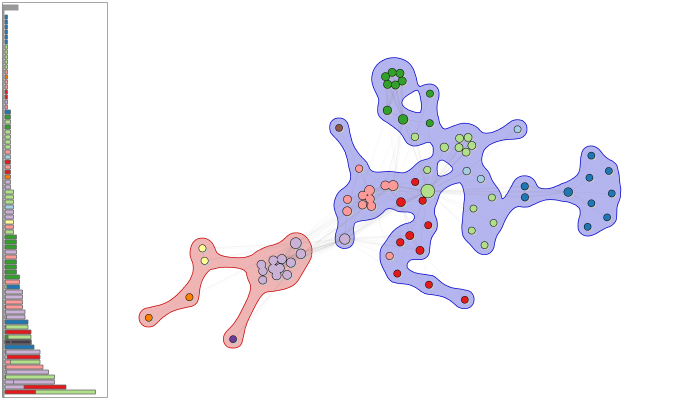}
        \end{minipage}
    }
    \hspace{2pt}
    
    \caption{Two scenarios for the L\'es Miserables graph are shown where repulsion may be considered by a user (top) and the result of it (bottom).}
    \label{fig:contract_ex}
\end{figure}

\paragraph{Preview Hovering.} To help users preview the impact of a bar's selection when the mouse hovers over a bar in the barcode, set visualization can be employed, such as bubble sets~\cite{collins2009bubble} or kelp diagrams~\cite{dinkla2012kelp}. We employ bubble sets on the \emph{subsets of nodes} to differentiate which nodes belong to which subset. \autoref{fig:contract_ex} shows examples of bubble sets employed on a dataset. The bubble sets demonstrate two examples of before and after the PH feature is selected for repulsion, respectively.

\paragraph{Hyperbolic Zoom for a Large Barcode.} The number of bars is equal to the number of nodes in the graph minus one. In order to scale the barcode appropriately for a large graph, a scrollbar with hyperbolic zoom is placed to the right of the bars. As the scrollbar moves, the focus of the hyperbolic zoom is modified to emphasize the associated bars. An example of this can be seen in our accompanying software.

\subsection{Typical Usage Session}
\label{sec.vis.session}

After a graph is loaded and any adjustments are made to standard force strengths, the user begins to explore the PH features of the graph. To start, the contraction forces are explored by slowly adjusting the threshold higher (to the right), which enables finding when the compactness reaches a desirable level. 

Next, the PH features are explored for repulsive forces. PH defines the elements with the highest persistence to be the ``most important''. Therefore, we typically start by looking for high persistence bars with a higher subset ratio and/or bars that split between different categories for labeled data. Hovering over a bar first informs the user if the partitioning may be interesting. Examples of interesting partitions include, but are not limited to, partitions that are mixed together (see \autoref{fig:contract_ex:a}) or a cluster of nodes already spatially co-located that the user would like to move away from the rest of the graph (see \autoref{fig:contract_ex:b}).  If the user finds the partition interesting, the repulsive force is enabled by clicking. Typically, we found a good layout was achieved by selecting 5-10 bars for repulsion, although this is by no means a limit. Usually, the entire process took no more than a few minutes to sufficiently investigate and fine-tune the graph layout.

\begin{table}[!bh]
    \caption{Quantitative Analysis of Results}
    \label{tab:quant}
    \includegraphics[width=\linewidth]{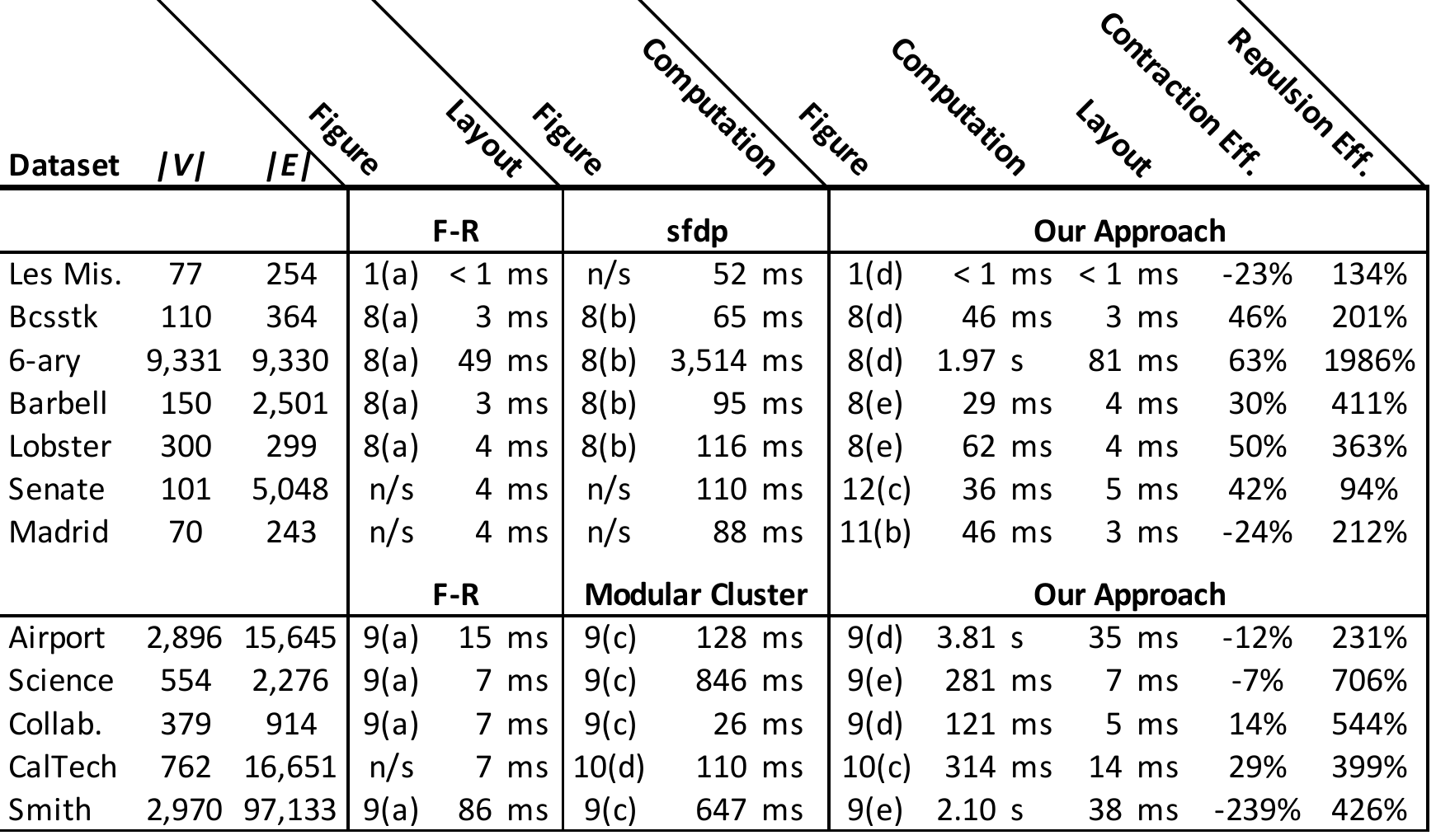}
    
    \vspace{-5pt}{ \footnotesize \hspace{5pt} n/s: Not shown}
    \vspace{-10pt}
\end{table}

\section{Results}

To evaluate using PH to guide force-directed layouts, we examine 12 datasets using hand-tuned layouts (see \autoref{tab:quant}). Our evaluation considers the following:
\vspace{-5pt}
\begin{itemize}
  \setlength{\itemsep}{2pt}
  \setlength{\parskip}{0pt}
  \setlength{\parsep}{0pt}
    \item We examine the scalability of our approach to quickly calculate PH and update the graph layout fast enough to support interactive visualization, using graphs up to 10,000 nodes or 100,000 edges.
    \item We measure the quality of the layouts produced by our approach. 
    \item We compare the layouts produced by our approach to popular force-directed layout methods and clustering techniques.
    \item We present three brief case studies on real-world data.
\end{itemize}

\subsection{Performance}

The accompanying software is implemented using Processing~\cite{processing}. All calculations are performed on the CPU. The force calculations are multithreaded. The live footage from the accompanying video was produced on a 2017 MacBook Pro with a 3.1 Ghz i5 processor to demonstrate the interactivity of the interface. 

Once the data is loaded, the $0$-dimensional PH features are extracted. The MST calculation is a variation of Kruskal's algorithm performed on the metric space. It is implemented using disjoint sets taking $O(|E| \alpha(|V|))$, where $\alpha$ is the inverse Ackermann function~\cite{cormen2009introduction}, an extremely slow growing function. After the MST is found, determining the node subsets (see \autoref{sec:ph:st}) takes $O(|V|)$.

For the F-R force-directed layout, we use the Barnes-Hut approximation~\cite{barnes1986hierarchical} for repulsive forces and standard pairwise springs. The total cost is $O(|V| \log |V| + |E|)$ per iteration. In addition, contraction forces are single pairwise springs. The worst case total cost is $O(|V|)$, if all PH features are selected for contraction. The repulsion force can be costly. Each force that is applied requires an additional run of the Barnes-Hut algorithm. Since the number of repulsive forces is usually small, the expected run time is $O(|V| \log |V|)$ with the worst case $O(|V|^2 \log |V|)$, when all PH features are set to repulse.

To improve interactivity of the visualization on larger datasets, certain noncritical features are disabled. For example, edge bundling is automatically disabled when the number of edges is greater than 500. In addition, bubble sets are disabled when the number of nodes is greater than 100. Instead, the graph nodes are surrounded by a halo, which is colored according to the set they belong to.

\newcommand{\placeFPS}[1]{\put(-20,0){
  		\begin{minipage}[t][0pt][t]{0pt}
			\tiny
			\mbox{FPS: #1}
		\end{minipage}
		}}
\newcommand{\placeDataset}[3]{\put(-105,90){
  		\begin{minipage}[t][0pt][t]{0pt}
			\tiny
			\mbox{#1} \\ \mbox{Nodes: #2} \\ \mbox{Edges: #3}
		\end{minipage}
		}}
		
\begin{figure*}[!t]
\centering
\begin{tabular}{ccccc}
    \hspace{15pt}
    \begin{minipage}[b]{0.170\linewidth}
		\vspace{2pt}
        \includegraphics[width=1.0\linewidth, height=34.0mm]{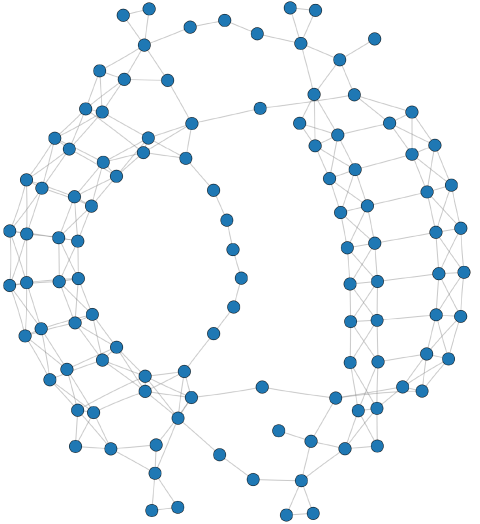}
	\end{minipage}
	\placeDataset{bcsstk}{110}{364}
	\placeFPS{60}

  & \includegraphics[width=0.170\linewidth, height=34.0mm]{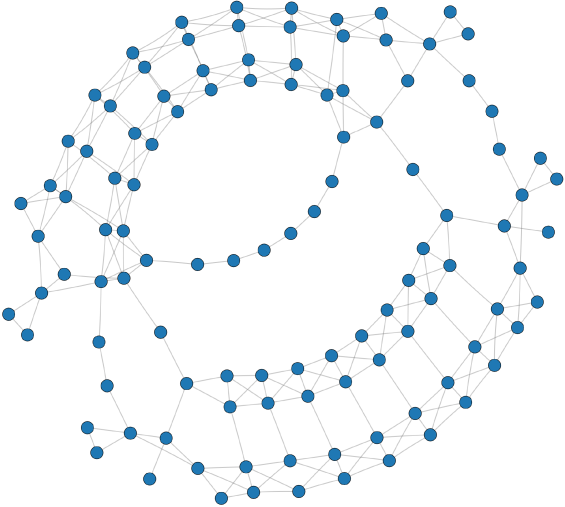}
      \placeFPS{n/a}
      
  & \includegraphics[width=0.170\linewidth, height=34.0mm]{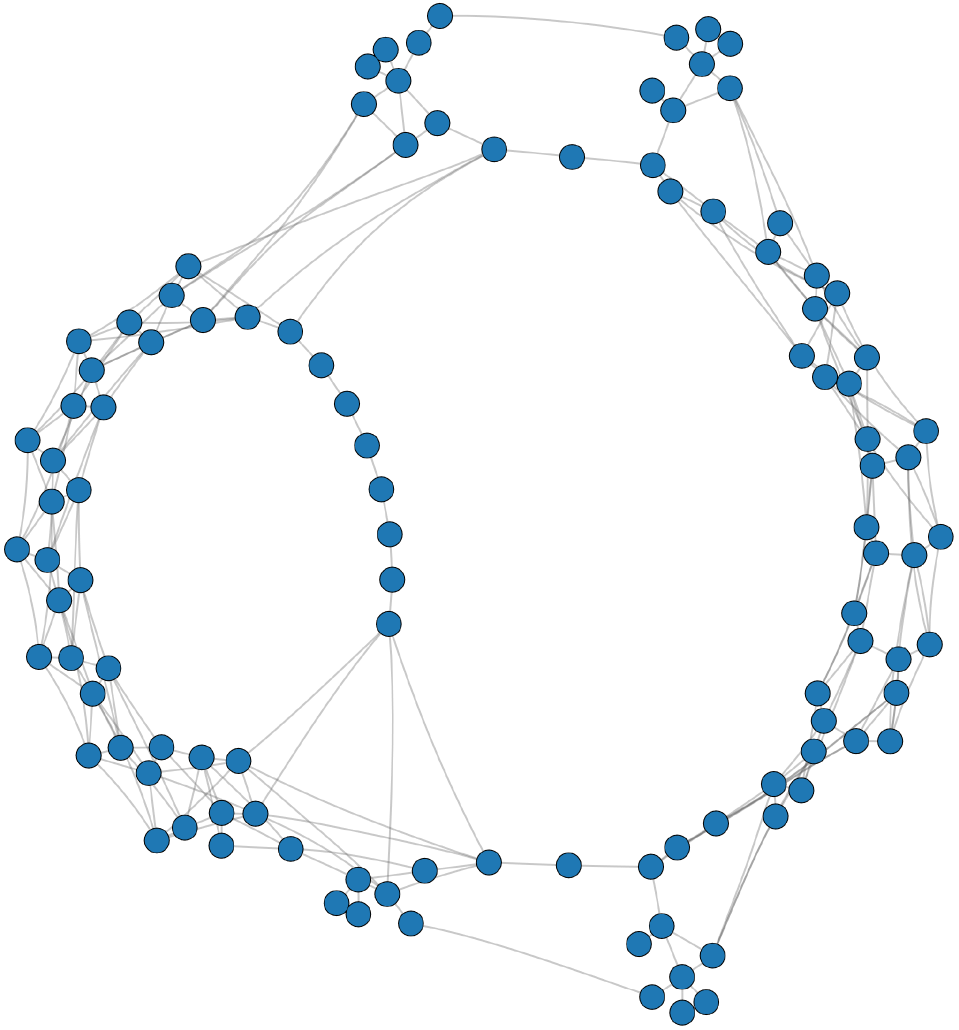}
    \placeFPS{60}
        
  & \includegraphics[width=0.170\linewidth, height=34.0mm]{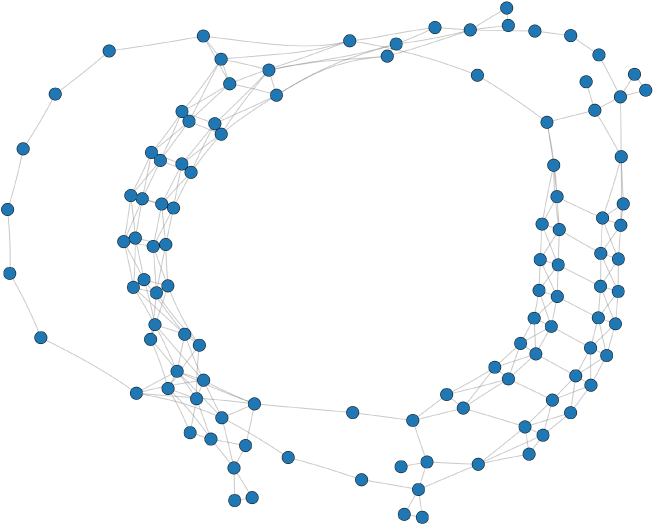}
  	\placeFPS{60}
  	  
  & \includegraphics[width=0.170\linewidth, height=34.0mm]{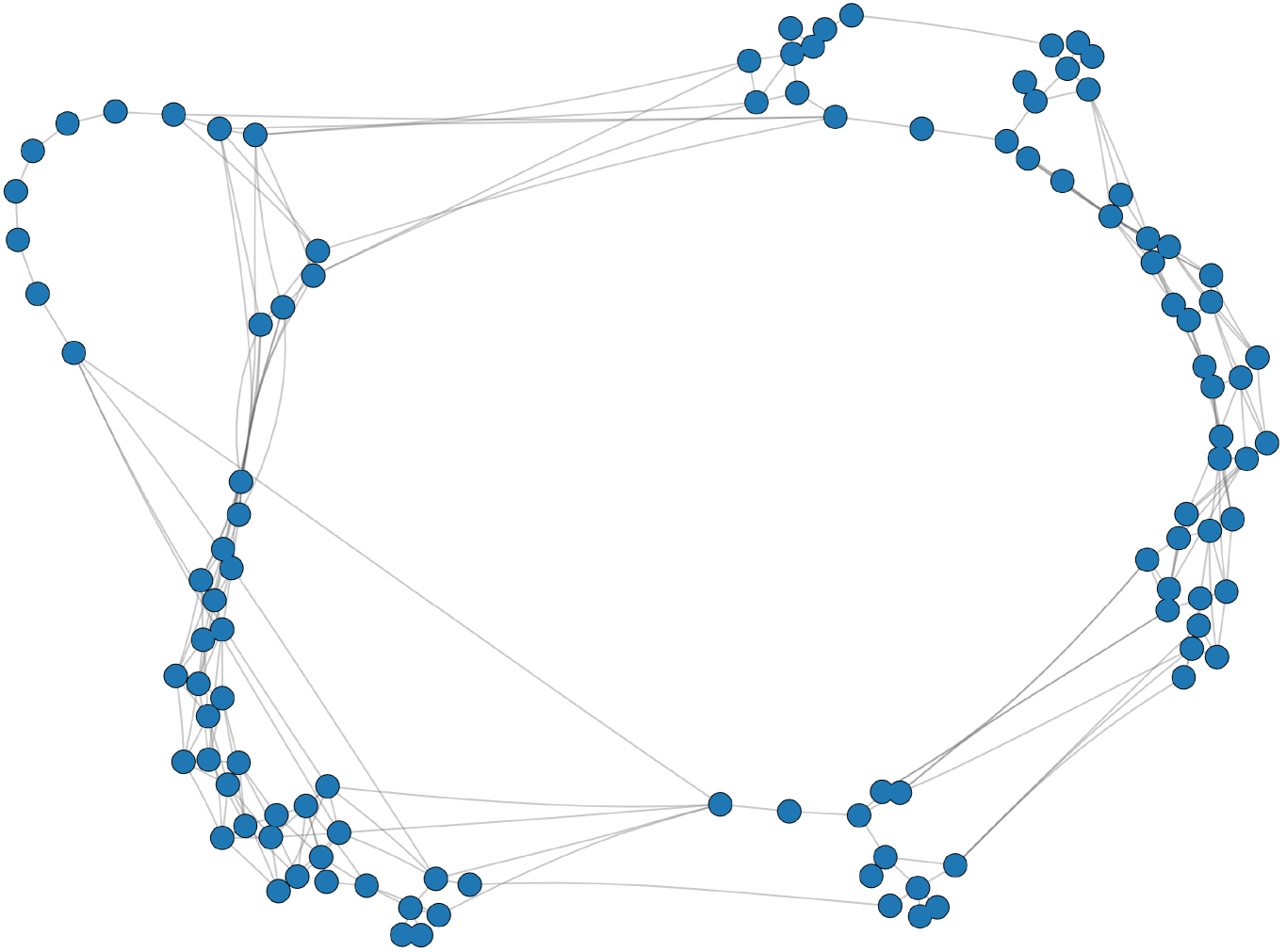} 
    \placeFPS{60} \\
  
    \hspace{15pt}
    	\begin{minipage}[b]{0.170\linewidth}
		\vspace{2pt}
        \includegraphics[width=1.0\linewidth, height=32.0mm]{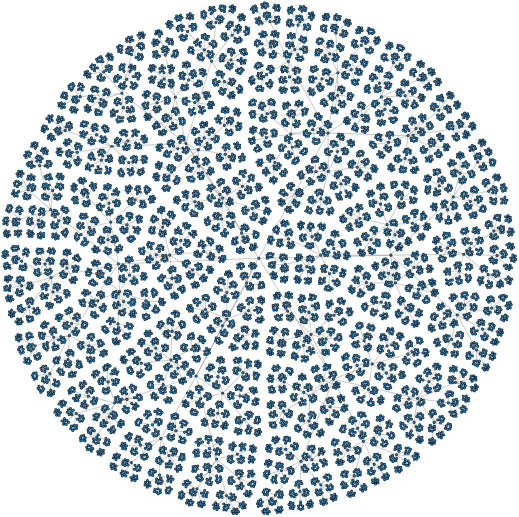}
		\end{minipage}
		\placeDataset{6-ary}{9331}{9330}
		\placeFPS{10}

  & \includegraphics[width=0.170\linewidth, height=32.0mm]{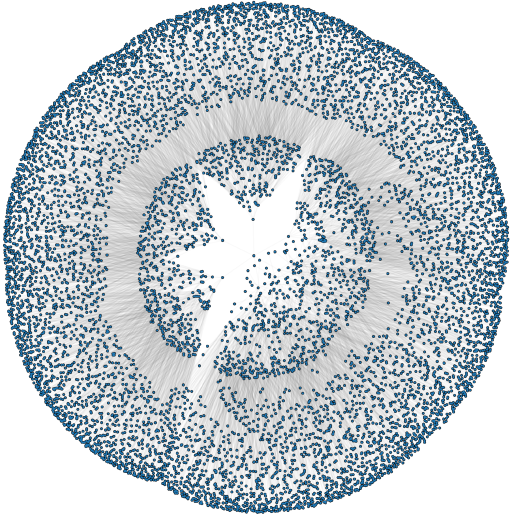}
    \placeFPS{n/a}
  
  & \includegraphics[width=0.170\linewidth, height=32.0mm]{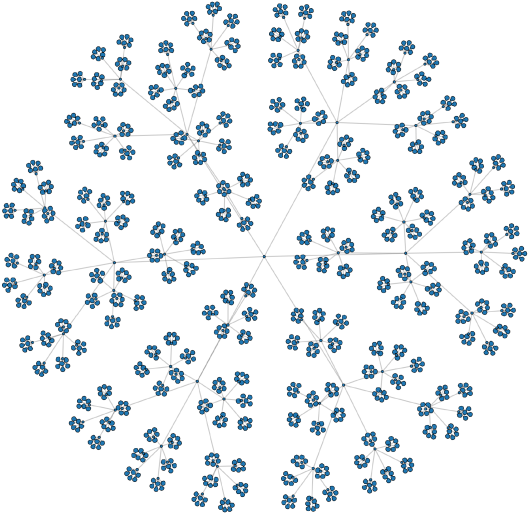} 
    \placeFPS{10}
  	    
  & \includegraphics[width=0.170\linewidth, height=32.0mm]{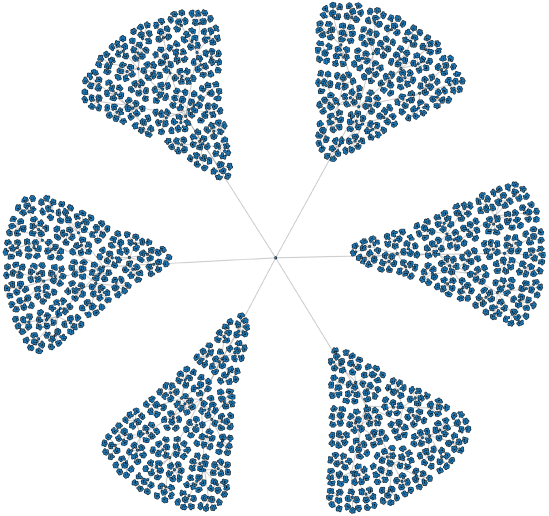}
    \placeFPS{7}
  	    
  & \includegraphics[width=0.170\linewidth, height=32.0mm]{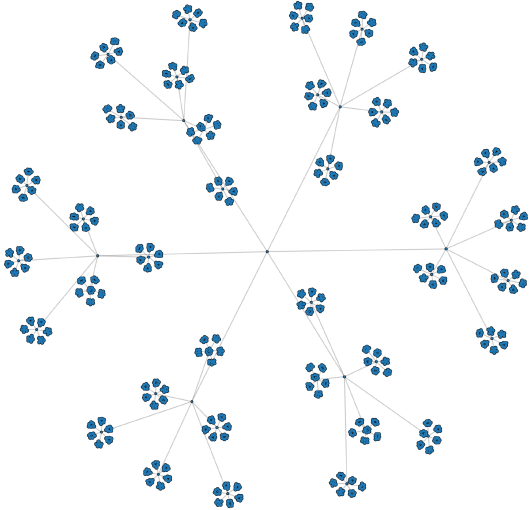} 
    \placeFPS{7} \\
  
    \hspace{15pt}
    \begin{minipage}[b]{0.170\linewidth}
        \vspace{2pt}
         \includegraphics[width=1.0\linewidth, height=34.0mm]{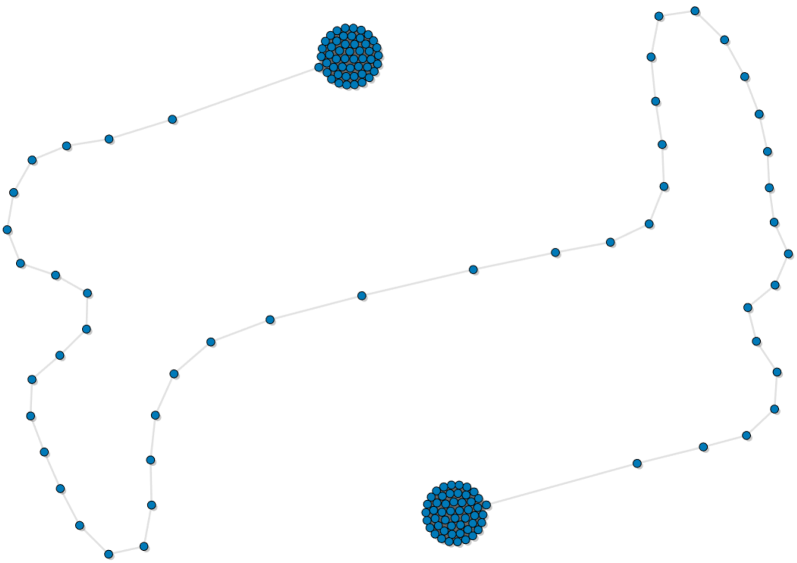} 
    \end{minipage}
    \placeDataset{Barbell}{150}{2501}
    \placeFPS{60}
		
  & \includegraphics[width=0.170\linewidth, height=34.0mm]{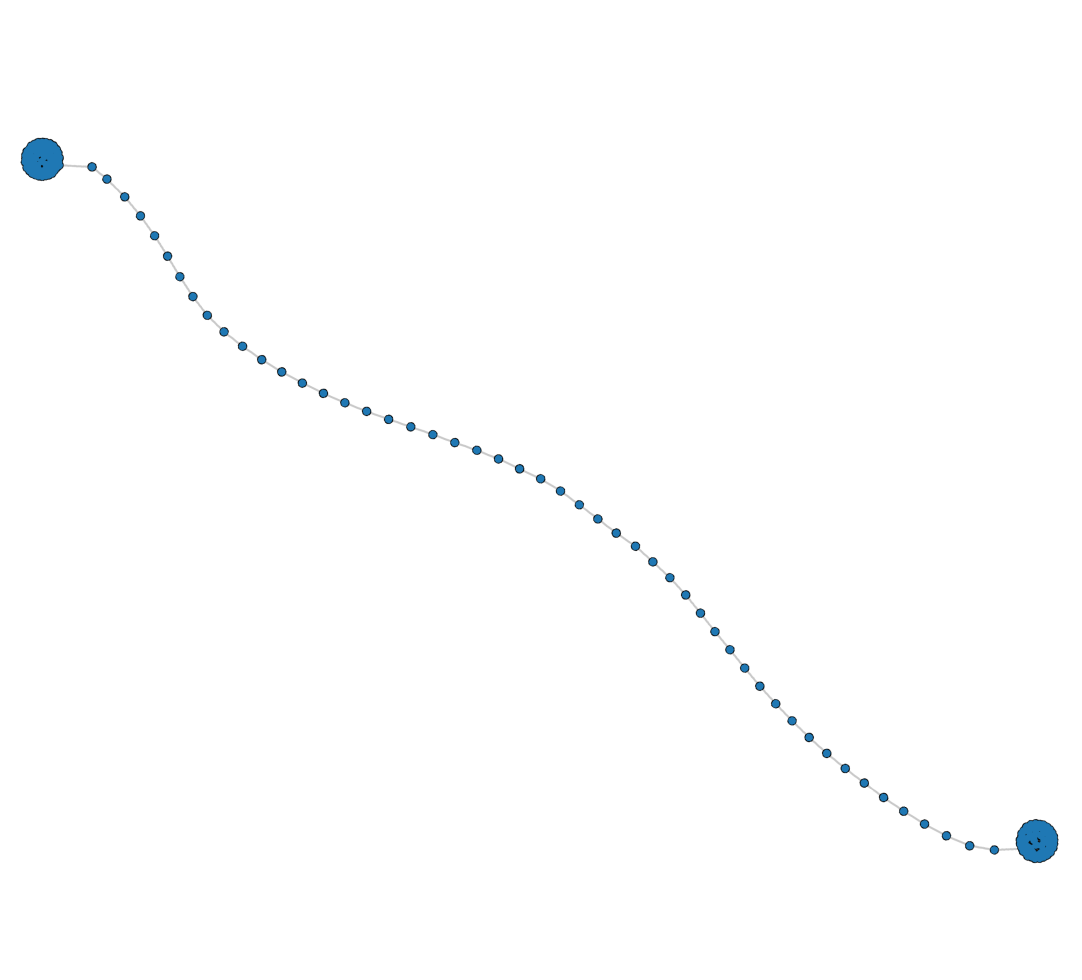} 
    \placeFPS{n/a}

  & \includegraphics[width=0.170\linewidth, height=34.0mm]{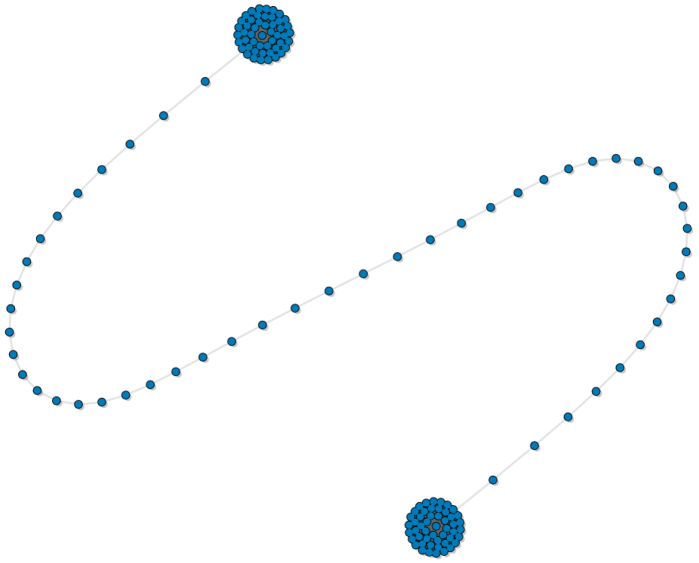}
    \placeFPS{60}
  	    
  & \includegraphics[width=0.170\linewidth, height=34.0mm]{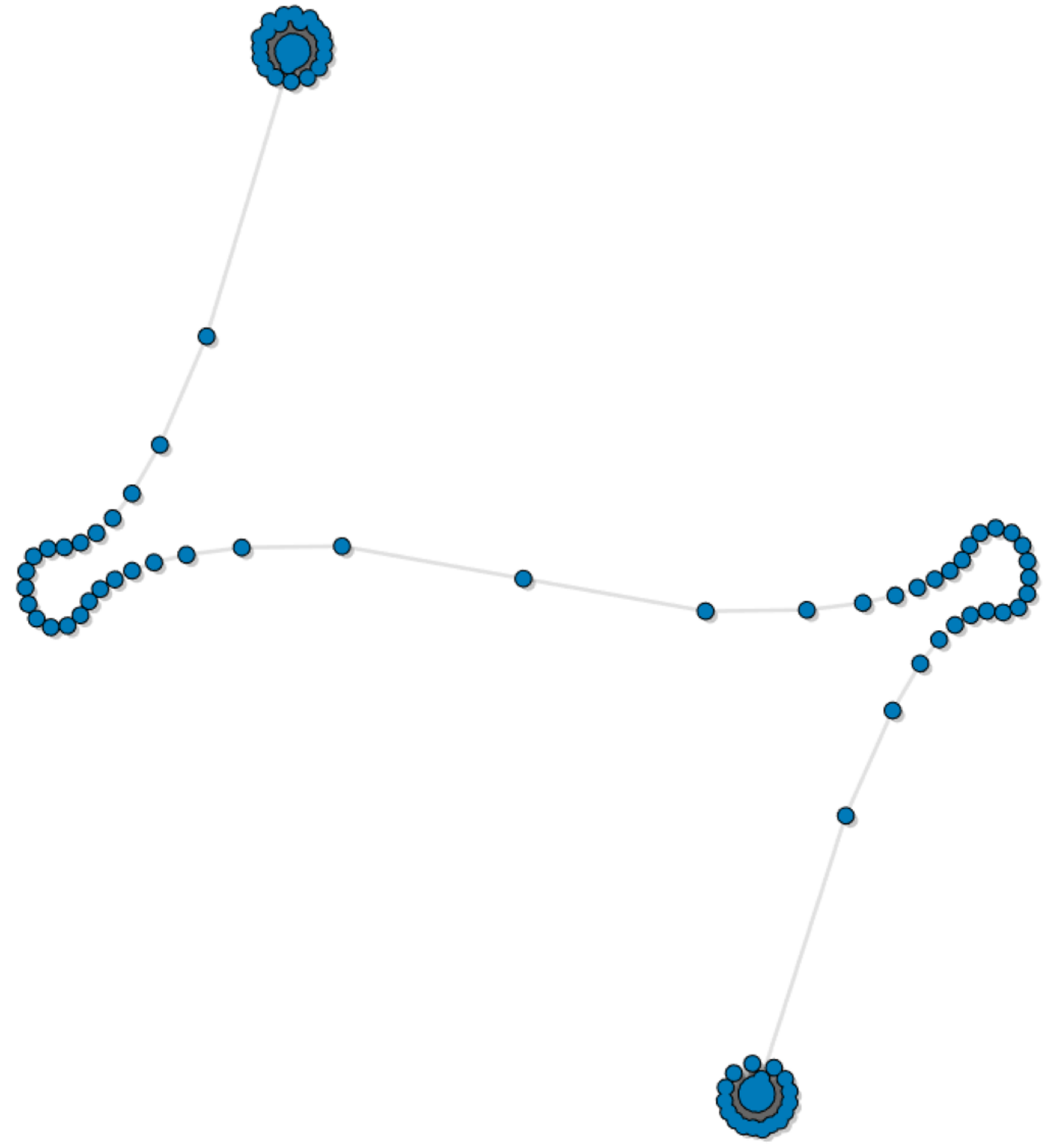}
    \placeFPS{60}
        
  & \includegraphics[width=0.170\linewidth, height=34.0mm]{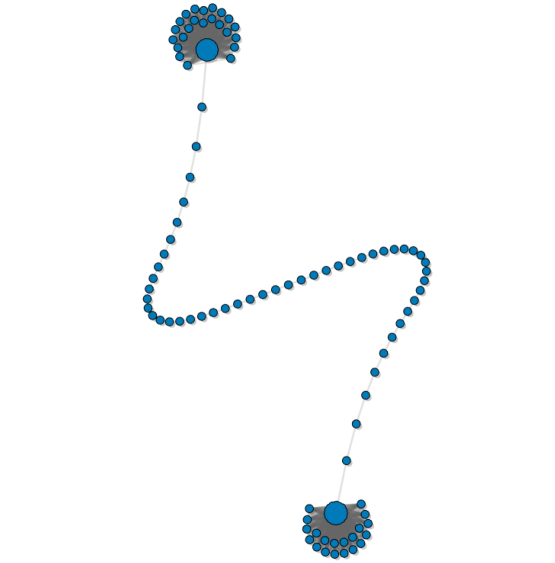} 
    \placeFPS{60} \\  
  	    
    \hspace{15pt}
    \begin{minipage}[b]{0.170\linewidth}
		\vspace{2pt}
        \includegraphics[width=1.0\linewidth, height=34.0mm]{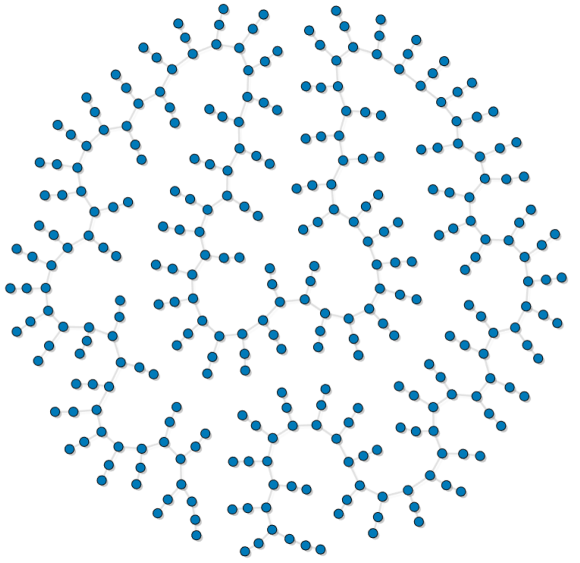} 
	\end{minipage}
	\placeDataset{Lobster}{300}{299}
	\placeFPS{60}

  & \includegraphics[width=0.170\linewidth, height=34.0mm]{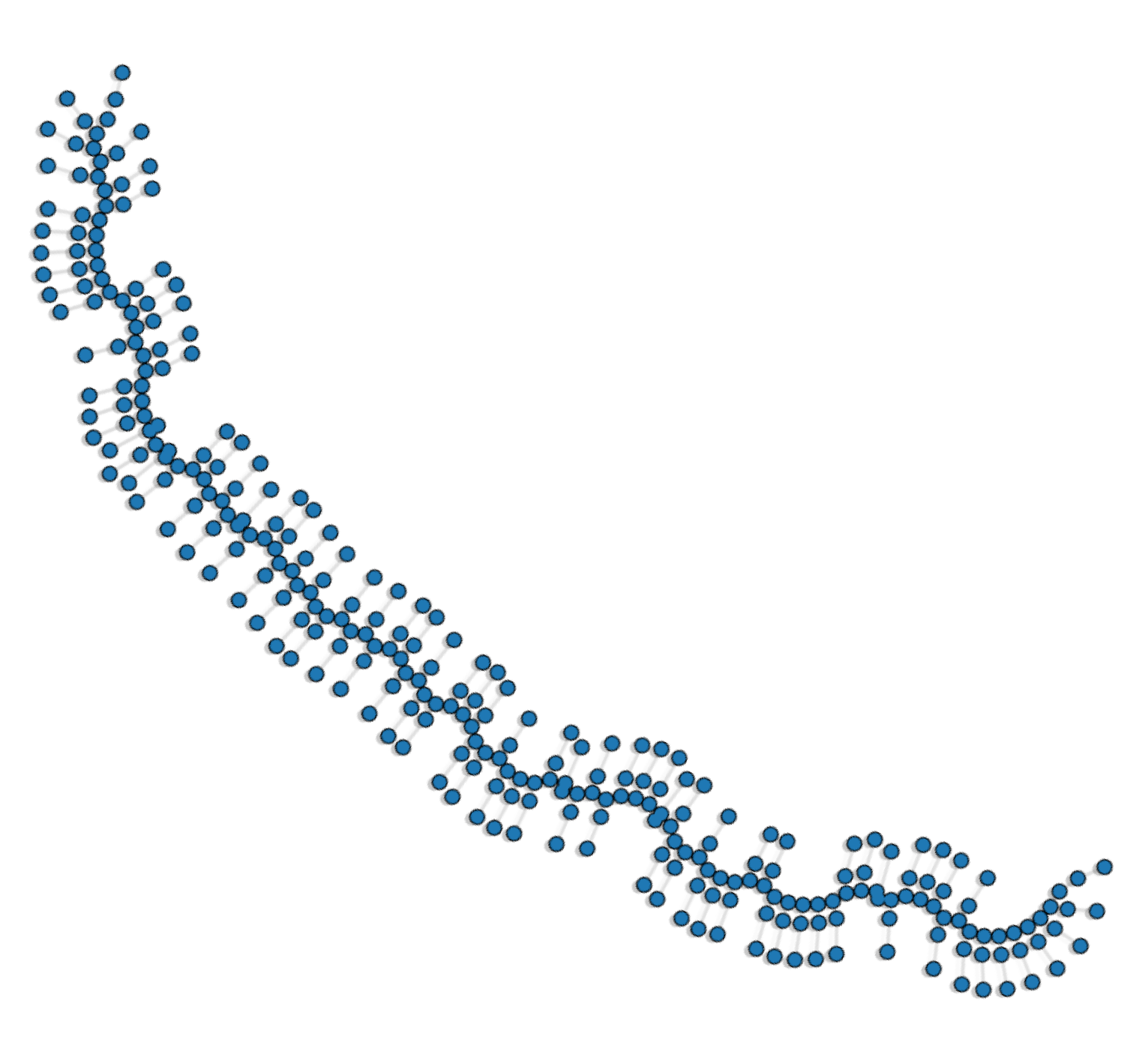} 
    \placeFPS{n/a}

  & \includegraphics[width=0.170\linewidth, height=34.0mm]{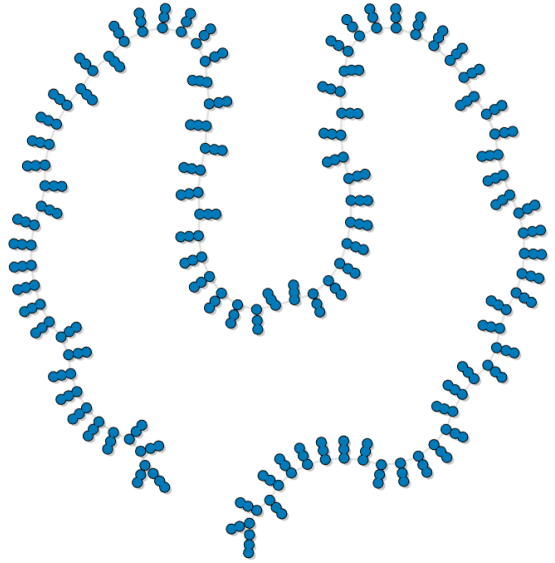}
    \placeFPS{60}
	    
  & \includegraphics[width=0.170\linewidth, height=34.0mm]{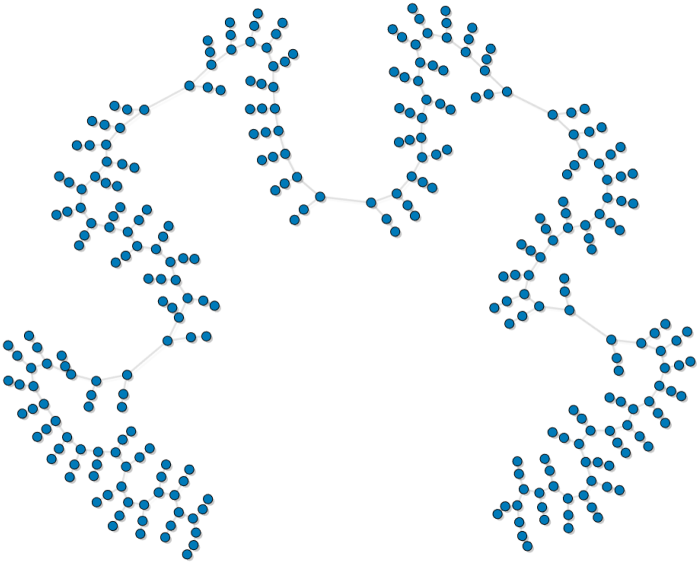}
    \placeFPS{60}
	    
  & \includegraphics[width=0.170\linewidth, height=34.0mm]{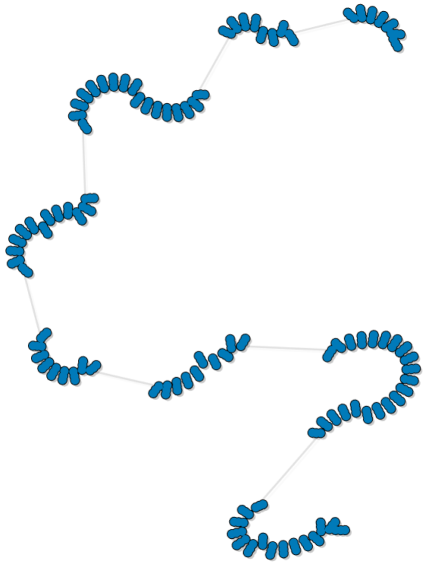} 
    \placeFPS{60} \\  
  
\end{tabular}

    \hfill
    \subfigure[F-R Layout]{\hspace{0.22\linewidth}}\hfill
    \subfigure[Neato Layout]{\hspace{0.185\linewidth}}\hfill
    \subfigure[Ours: Contracted]{\hspace{0.185\linewidth}}\hfill
    \subfigure[Ours: Repulsed]{\hspace{0.185\linewidth}}\hfill
    \subfigure[Ours: Combination]{\hspace{0.185\linewidth}}\hfill
    
    \caption{Illustration of our approach on the synthetic graph examples: (a) a Fruchterman-Reingold (F-R) force-directed layout, hand-tuned; (b) a Neato layout~\cite{north2004drawing} generated by Graphviz~\cite{ellson2002graphviz} using Jaccard index for edge weights; (c) our approach: only contraction is applied; (d) our approach: only repulsion is applied; and (e) our approach: both contraction and repulsion are applied.}
    \label{results.synthetic}
\end{figure*}

\newcommand{\placeClusterN}[1]{\put(-25,6){
  		\begin{minipage}[t][0pt][t]{0pt}
			\tiny
			\mbox{Clusters: #1}
		\end{minipage}
		}}

\begin{figure*}[!t]
\centering
\begin{tabular}{ccccc}
 	\hspace{30pt}
	\begin{minipage}[b]{0.165\linewidth}
		\vspace{2pt}
    	\includegraphics[width=1.0\linewidth,height=32mm]{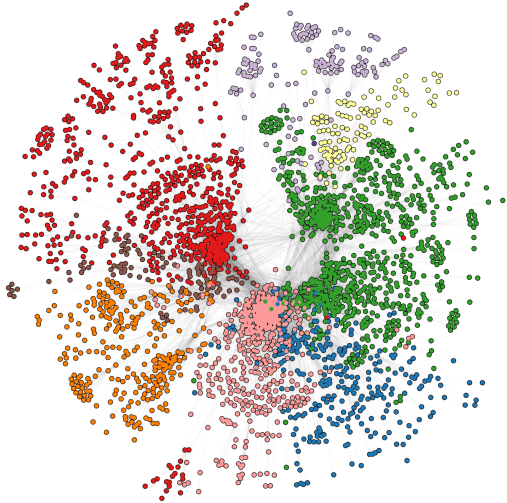}
    \end{minipage}
    \put(-20,-5){\placeDataset{Airport}{2896}{15645}}
	\placeFPS{15}
      \put(-125,8){
			\includegraphics[width=.096in]
            {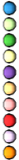}
            }
      \put(-115,2){
  		\begin{minipage}[t][0pt][b]{0pt}
        	\tiny
            Africa \\
			Antarctica \\
			Asia \\
            Australasian \\
            Australia \\
            \mbox{C.\ America} \\
			Europe \\
			\mbox{N.\ America} \\
            Oceania \\
			\mbox{S.\ America} \\
		\end{minipage}
		}      

 & \includegraphics[width=0.165\linewidth, height=32mm]{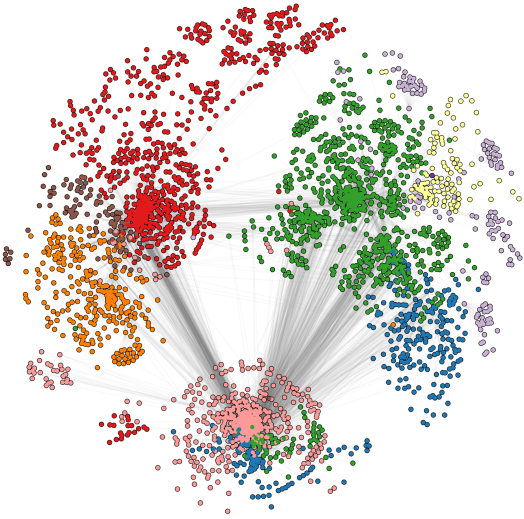}    	  
		\placeFPS{15}
		\placeClusterN{3}

 & \includegraphics[width=0.165\linewidth, height=32mm]{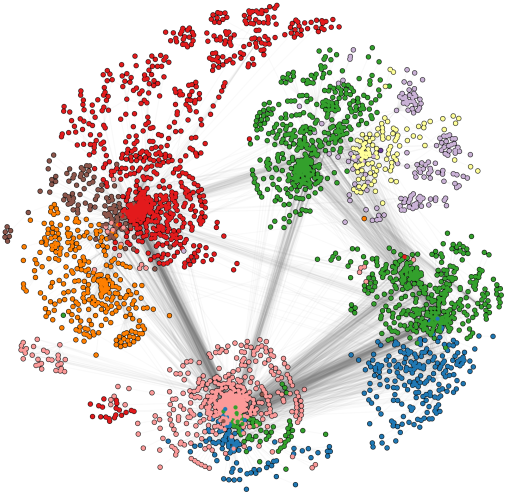}    	  
		\placeFPS{15}
		\placeClusterN{4}
		
 & \includegraphics[width=0.165\linewidth, height=32mm]{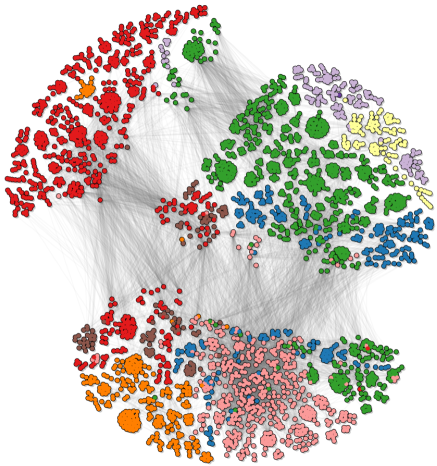} 
		\placeFPS{15}

 & \includegraphics[width=0.165\linewidth, height=32mm]{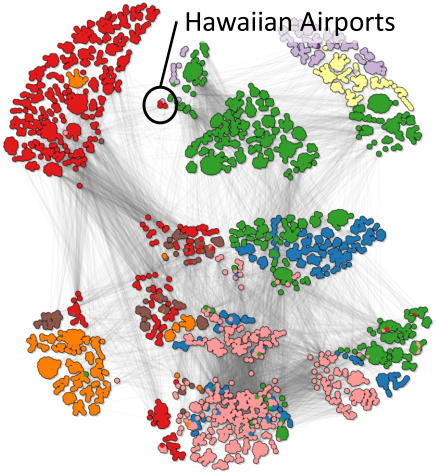}   	  
		\placeFPS{15}

	\\ 
        
 	\hspace{30pt}
	\begin{minipage}[b]{0.165\linewidth}
		\vspace{2pt}
    	\includegraphics[width=1.0\linewidth, height=32mm]{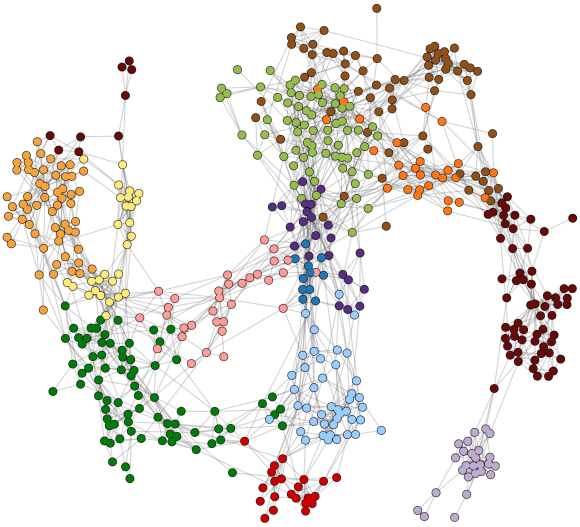}
    \end{minipage}
    \put(-20,-3){\placeDataset{Science}{554}{2276}}
    \placeFPS{60}
      \put(-125,0){
		\includegraphics[trim=3pt 0 0 0, clip, width=.0785in]{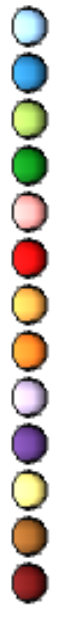}}
      \put(-118,2){
  		\begin{minipage}[t][0pt][b]{0pt}
        	\tiny
			\mbox{Biology} \\
			\mbox{Biotech} \\
			\mbox{Med.\ Spec.} \\
			\mbox{Che/Mec/Civ} \\
			\mbox{Chem.} \\
			\mbox{Earth Sci.} \\
			\mbox{EE/CS} \\
			\mbox{Brain Res.} \\
			\mbox{Humanities} \\
			\mbox{Math/Phys.} \\
			\mbox{Health Prof.} \\
			\mbox{Social Sci.}
		\end{minipage}
		}             
 & \includegraphics[width=0.165\linewidth, height=32mm]{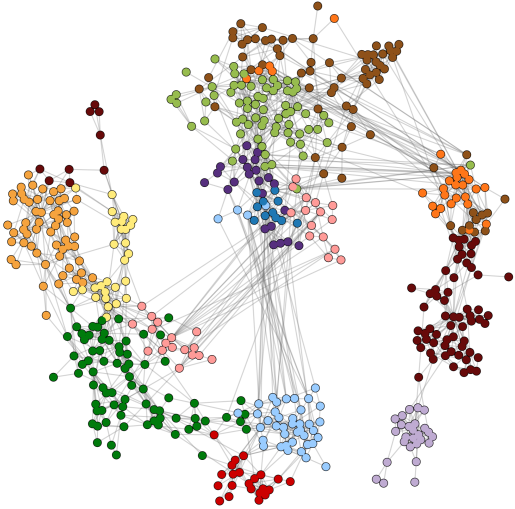}
   	  \placeFPS{60}
   	  \placeClusterN{3}
   	  
 & \includegraphics[width=0.165\linewidth, height=32mm]{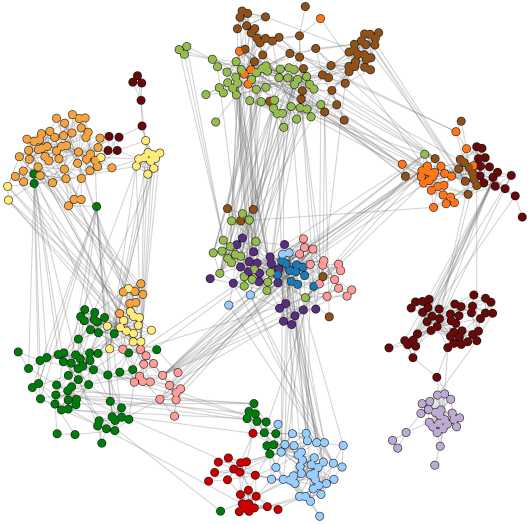}
   	  \placeFPS{60}
   	  \placeClusterN{7}
   	  
  & \includegraphics[width=0.165\linewidth, height=32mm]{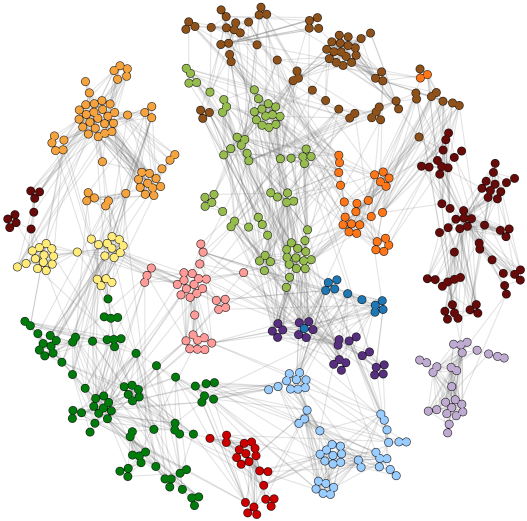} 
 			\placeFPS{60}
& \includegraphics[width=0.165\linewidth, height=32mm]{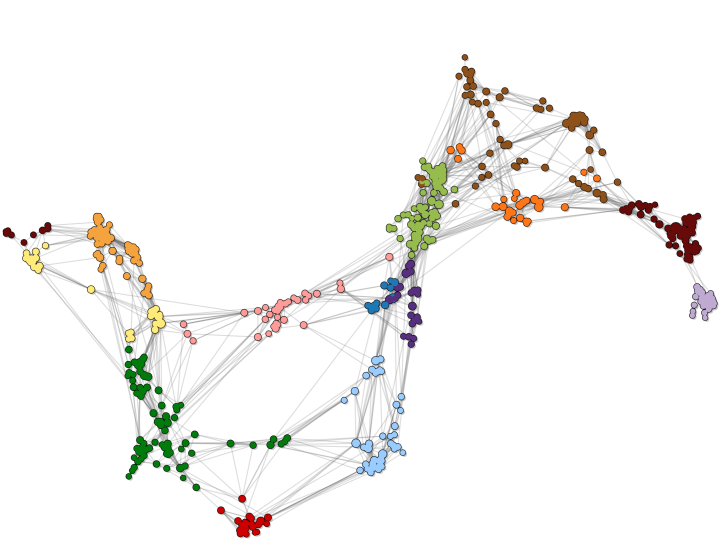} 
			\placeFPS{60}

	\\  
 
 	\hspace{30pt}
	\begin{minipage}[b]{0.165\linewidth}
		\vspace{2pt}
    	\includegraphics[width=1.0\linewidth, height=32mm]{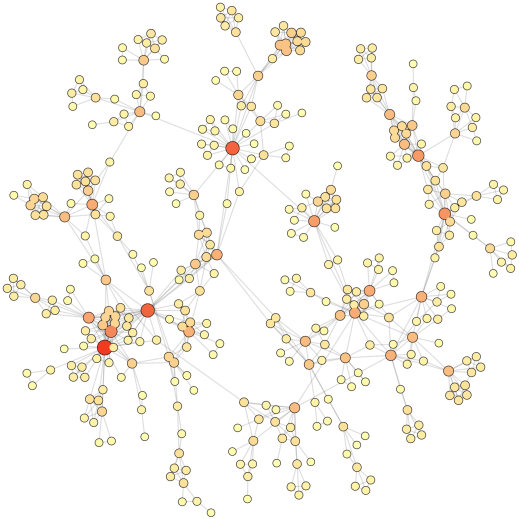} 
    \end{minipage}
    \put(-20,-5){\placeDataset{Collaboration}{379}{914}}
    \placeFPS{60}
      \put(-125,0){
  		\begin{minipage}[t][0pt][t]{0pt}
			\tiny
			\includegraphics[width=.10in]{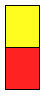}
		\end{minipage}
		}  
        
        \put(-115,3){
  		\begin{minipage}[t][0pt][t]{0pt}
			\tiny
			\mbox{High degree}
		\end{minipage}
		}
        
        \put(-115,10){
  		\begin{minipage}[t][0pt][t]{0pt}
			\tiny
			\mbox{Low degree}
		\end{minipage}
		}
        
 & \includegraphics[width=0.165\linewidth, height=32mm]{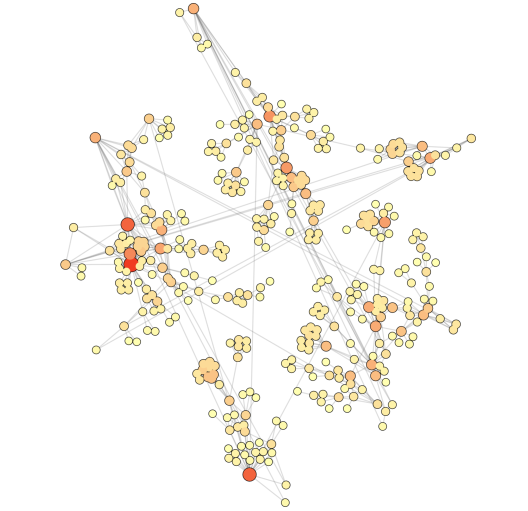}
   	  \placeFPS{60}
   	  \placeClusterN{5}
 & \includegraphics[width=0.165\linewidth, height=32mm]{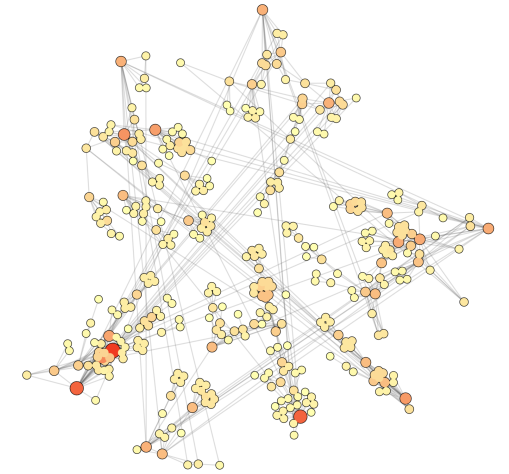}
   	  \placeFPS{60}
   	  \put(-3,0){\placeClusterN{10}}
   	  
 & \includegraphics[width=0.165\linewidth, height=32mm]{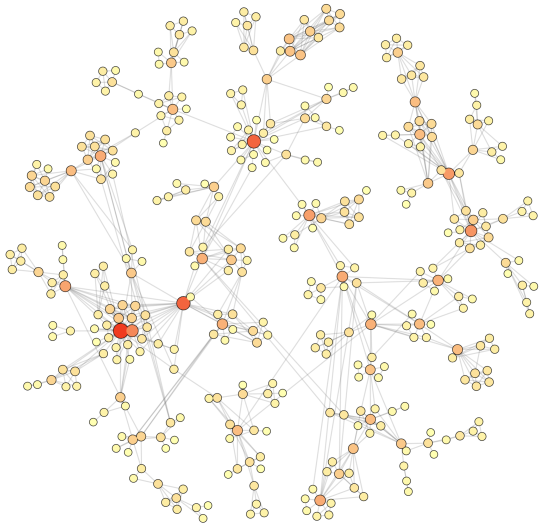}
			\placeFPS{60}
 & \includegraphics[width=0.165\linewidth, height=32mm]{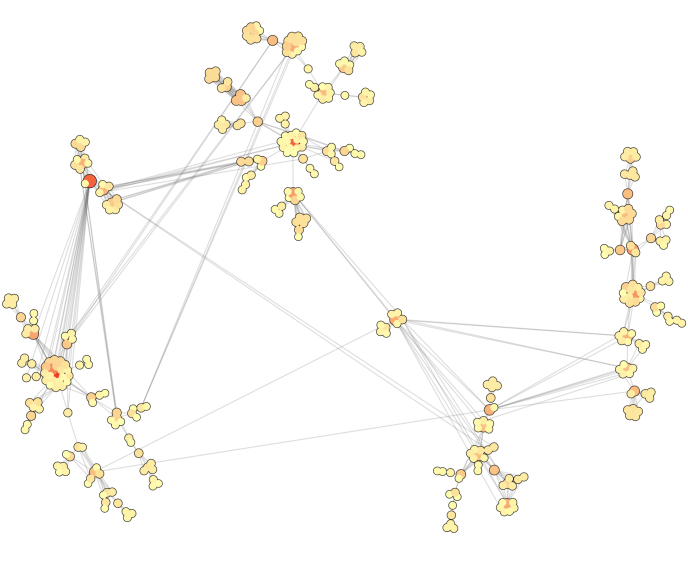} 
			\placeFPS{60}
		\\

 	\hspace{30pt}
	\begin{minipage}[b]{0.165\linewidth}
		\vspace{2pt}
        \includegraphics[width=1.0\linewidth, height=32mm]{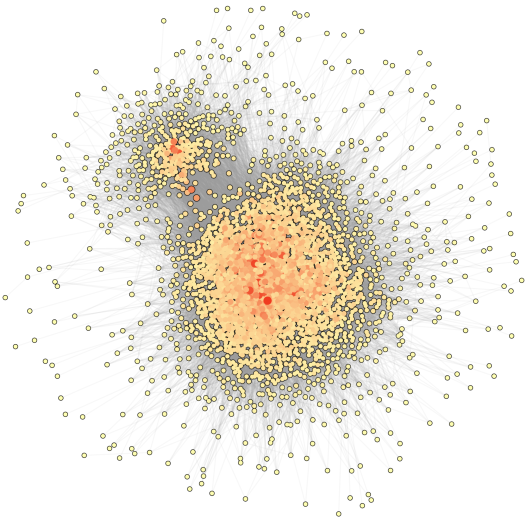} 
    \end{minipage}
    \put(-20,-5){\placeDataset{Smith}{2970}{97133}}
    \placeFPS{15}
      \put(-125,0){
  		\begin{minipage}[t][0pt][t]{0pt}
			\tiny
			\includegraphics[width=.10in]{figures/legend_degree.png}
		\end{minipage}
		}  
        
      \put(-115,3){
  		\begin{minipage}[t][0pt][t]{0pt}
			\tiny
			\mbox{High degree}
		\end{minipage}
		}
        
      \put(-115,10){
  		\begin{minipage}[t][0pt][t]{0pt}
			\tiny
			\mbox{Low degree}
		\end{minipage}
		}

 & \includegraphics[width=0.165\linewidth, height=32mm]{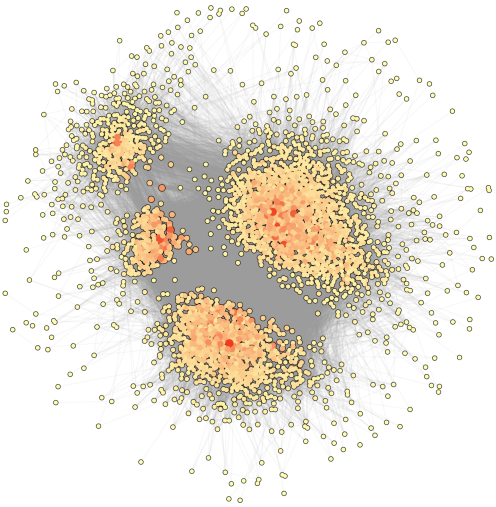}
   	  \placeFPS{15}
   	  \placeClusterN{3}
   	  
 & \includegraphics[width=0.165\linewidth, height=32mm]{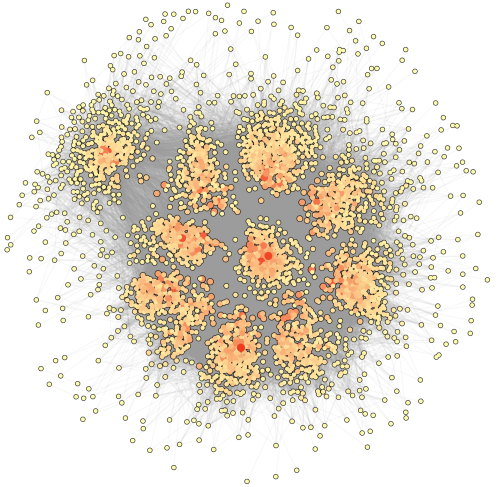} 
   	  \placeFPS{15}
   	  \put(-3,0){\placeClusterN{11}}

& \includegraphics[width=0.15\linewidth, height=32mm]{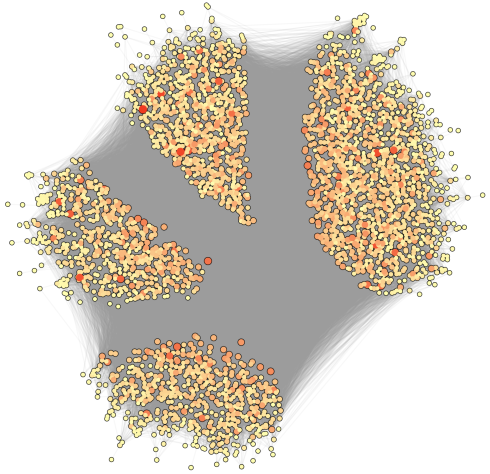} 
 			\placeFPS{15}
& \includegraphics[width=0.15\linewidth, height=32mm]{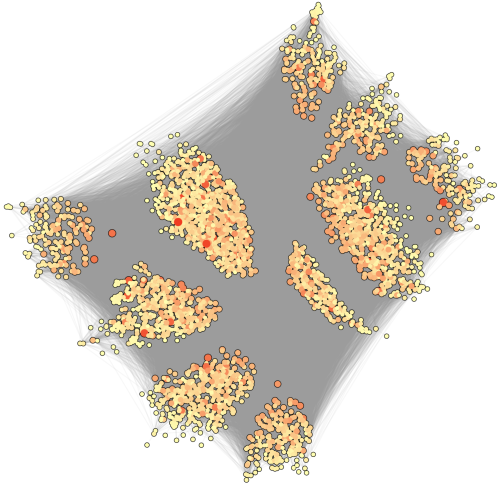}  
			\placeFPS{10}
\end{tabular}

    \vspace{-5pt}
    \hspace{28pt}
    \subfigure[F-R Layout]{\hspace{0.185\linewidth}}\hfill
    \subfigure[MH Clustering Ex 1]{\hspace{0.185\linewidth}}\hfill
    \subfigure[MH Clustering Ex 2]{\hspace{0.185\linewidth}}\hfill
    \subfigure[Ours: Ex 1]{\hspace{0.185\linewidth}}\hfill
    \subfigure[Ours: Ex 2]{\hspace{0.185\linewidth}}\hfill
    
    \caption{Illustration of our approach on a series of graph examples: (a) Fruchterman-Reingold force-directed layout, hand-tuned; (b-c) 2 examples of modularity hierarchical clustering with cluster number selected manually; (d-e) 2 examples using our approach.}
    \label{results.real}
\end{figure*}

To show that the performance of our approach is \textit{comparable} to other techniques, we report computation times in \autoref{tab:quant}. For our approach, the category \textit{Computation} is the time needed to calculate the PH and determine the node subsets. For Neato~\cite{north2004drawing}, it is the time to converge. For hierarchical clustering, it is the time to compute the entire hierarchy. These computations occur only once, when the data is loaded. The category \textit{Layout} is the time in milliseconds (ms) needed for one iteration of the force-directed layout calculation. Our implementation runs one iteration per rendering frame. In addition, many of our examples include frame rate, reported as frames per second (FPS). This number is generally less relevant, as it includes the extra costs of rendering, edge bundling, etc., which are mostly fixed costs. 

The results demonstrate that, for all datasets, our approach has the scalability necessary to be utilized in interactive visualization. The PH calculations take at most a few seconds, and the time required for most layouts is less than 10 ms; for the larger graphs tested, it is  always less than 100 ms.

\subsection{Layout Quality}
\label{sec.results.layoutQ}

To assess the responsiveness of our approach, we introduce a measure specifically designed to quantify how well the resulting layouts reflect user intentions in emphasizing or de-emphasizing selected PH features. The approach works by comparing the PH of the embedded graph nodes of the layout before and after user modification. The approach starts by calculating the PH of the nodes of the \textit{source} F-R force-directed layout (without any contraction or repulsion) using Euclidean distance between nodes\footnote{Note, this calculation does \textit{not} consider the connectivity of the graph.}. Then, the PH of the user-selected \textit{target} layout is calculated similarly.

Given a source and target, we extract the set of bars from each that are  selected by the user for contraction ($C$) and repulsion ($R$). The persistence of those bars in the source and target is $P_S$ and $P_R$, respectively. The effect of contraction ($E_C$) and repulsion ($E_R$) are calculated as:

\begin{minipage}{0.45\linewidth}
$$ E_C =  \frac{1}{|C|} \sum_{x \in C} \frac{P_S(x) - P_T(x)}{P_S(x)}$$
\end{minipage}, 
\begin{minipage}{0.45\linewidth}
$$ E_R =  \frac{1}{|R|} \sum_{x \in R} \frac{P_T(x)-P_S(x)}{P_S(x)} $$
\end{minipage}
\vspace{5pt}

The results appear in the final two columns of \autoref{tab:quant}, comparing the source layout (F-R layout listed in the table) to the target layout (our approach). Intuitively, this measure quantifies how much on average the features of a layout have been contracted or repulsed in the target layout relative to the source layout---negative is undesirable; zero means no impact; positive is desirable and the larger the better. 

For the most part, our approach shows a substantial positive impact from the applied contraction and repulsion, indicating the user's intentions have been reflected in the layout. The notable exceptions are that some datasets have a negative contraction effect. For these examples, negativity does not mean that the contraction is entirely ineffective---in reality, the repulsive effect just \emph{overpowers} the contractive effects on some parts of the layout, leading to an average effect that is negative.

\subsection{Comparison to Other Techniques}

\subsubsection{Comparison to Popular Force-Directed Layout Methods}

We test our method on four synthetic unweighted graphs in \autoref{results.synthetic} to compare the layouts of F-R force-directed layout, Neato~\cite{north2004drawing} from Graphviz~\cite{ellson2002graphviz}, and our approach.  \emph{Bcsstk} is taken from the UF Sparse Matrix Collection~\cite{davis2011university}; \emph{6-ary}, \emph{Barbell}, and \emph{Lobster} are generated using NetworkX~\cite{hagberg2008exploring}. \emph{Bcsstk} is a symmetric stiffness graph containing $110$ nodes and $364$ edges. \emph{6-ary} is a balanced tree of depth five containing $9331$ nodes and $9330$ edges. \emph{Barbell} is a simple graph connecting two complete subgraphs of $50$ nodes each with a bridge of $50$ nodes, totaling $150$ nodes and $2501$ edges. \emph{Lobster} is a tree with the property that the removal of leaves results in a caterpillar graph~\cite{golomb1996polyominoes}. For all layout methods, each graph has weights applied by using the method described in Sect.~\ref{sec:phUnweighted} with the following neighborhood size: {Bcsstk}: 1-hop; \emph{6-ary}: 2-hop; \emph{Barbell}: 1-hop; and \emph{Lobster}: 1-hop. \autoref{results.synthetic} shows three examples of our approach using contraction only, repulsion only, and a combination of both.

\subsubsection{Comparison to Hierarchical Clustering}

Next, we compare our approach to modularity-based hierarchical clustering. For a survey of graph clustering, including modularity, see~\cite{schaeffer2007graph}.

We use a greedy approach to form the clusters~\cite{newman2003structure,noack2009multi} available in Graphviz~\cite{ellson2002graphviz}, since the optimal version is NP-hard~\cite{clemenccon2012hierarchical}. The algorithm begins by initializing each node in the graph into its own cluster. The 2 clusters whose merging will cause the largest increase in modularity are then combined. The weighted modularity is calculated as~\cite{lou2011finding}:
$$Q_w = \frac{1}{2w_s} \sum_{ij}\left[ w_{ij} - \frac{w_i w_j}{2w_s} \right] \delta_{c_i,c_j},$$ 
where $2w_s= \sum_{ij}w_{ij}$, $w_i=\sum_j w_{ij}$, and Kronecker delta $\delta_{c_i,c_j}$ is $1$ if both $i$ and $j$ are in the same community, $0$ otherwise. This process is repeated until a single cluster remains. 

To reflect the clustering in the graph layout, we increase the spring resting lengths between nodes of different clusters. Examples can be seen in \autoref{results.real}. \autoref{tab:quant} shows the time to compute the clustering.

\begin{figure}[!ttb]
    \centering

    \subfigure[Adaptive Refinement from~\cite{nocaj2016adaptive}]{\includegraphics[height=1.2in]{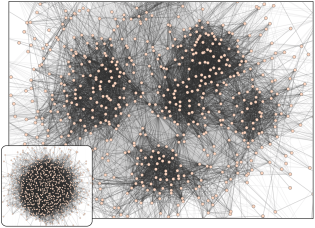}}
    \subfigure[Adaptive Refinement from~\cite{nocaj2014untangling}]{\includegraphics[height=1.2in]{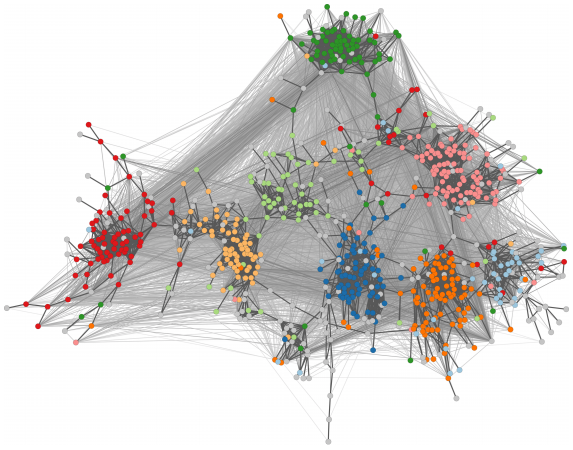}}    
    
    \vspace{-2pt}
    \subfigure[Modularity Hierarchical Clustering]{\hspace{8pt}\includegraphics[height=1.5in]{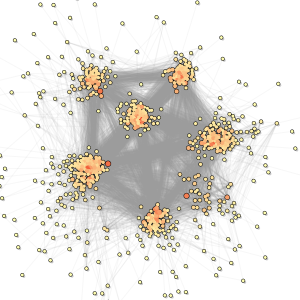}\hspace{8pt}}
    \subfigure[Our Approach]{\includegraphics[height=1.5in]{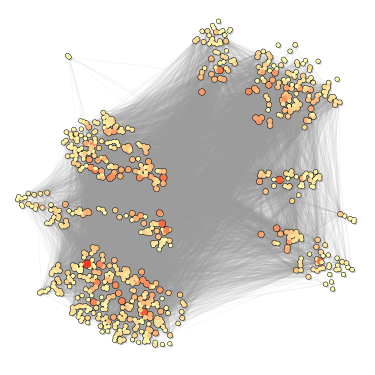}}

    \caption{Caltech datasets containing $762$ nodes and $16,651$ edges are compared using (a,b) 2 adaptive refinement techniques, (c) modularity hierarchical clustering, and (d) our approach.}
    \label{fig:caltech}
    \vspace{-6pt}
\end{figure}

\paragraph{2011 International Airports (Airport)} (\autoref{results.real} row 1) is taken from Openflights.org, where each node is an international airport, labeled by continental region, and edges are weighted by the number of routes between two airports. The largest connected component from this dataset contains $2,896$ nodes and $15,645$ edges. Using a combination of contraction and repulsion forces, \autoref{results.real} columns 4 and 5 reveal a split among Western, Eastern, and Central airports. One notable cluster is formed at the split of North America and Asia, containing several Hawaiian airports. The results in \autoref{results.real} columns 2 and 3 show similar clustering insights, but Hawaiian airports are not directly visible.

\paragraph{UCSD Map of Science (Science)} (\autoref{results.real} row 2)~\cite{borner2012design} is a map of $554$ subdisciplines in science, represented as nodes, and cross-disciplinary coauthorship as the $2,276$ edges. Our method, when applied to this particular dataset, results in a graph that retains the overall shape of the F-R layout with the addition of clustering communities that share similar disciplines. On the other hand, the clustered versions of the graph (columns 2 and 3) highlight the clustering structure but lose the context (ground truth labels).

\paragraph{The Collaboration Science Network (Collaboration)} (\autoref{results.real} row 3) \cite{collabData} is a coauthorship network with $379$ nodes--publishing scientists in network theory--and $914$ edges--a connection of two authors appearing on the same paper. The emphasis of the graph is to identify communities of collaborators. With the majority of bars selected for contraction (column 4 and 5), subcommittees are brought more tightly together, better revealing the overall graph shape. Clustering on the other hand (columns 2 and 3) produces results that we found difficult to interpret.

\paragraph{Smith College (Smith)} (\autoref{results.real} row 4) is from the Facebook100 dataset~\cite{traud2011comparing} and shows the social relations of students at Smith College. The graph has $2,970$ nodes and $97,133$ edges. The original graph contained attributes, including dormitory, gender, etc., but we were unable to locate this data. For this example, the key emphasis is on the scalability of our approach (see \autoref{tab:quant}).

\paragraph{Caltech} (\autoref{fig:caltech}) is a dataset of the social links at the California Institute of Technology, also from the Facebook100 dataset~\cite{traud2011comparing}. The graph has $762$ nodes and $16,651$ edges. For this example, we compare against modularity hierarchical clustering and adaptive refinement techniques~\cite{nocaj2016adaptive,nocaj2014untangling}. Each of the techniques presents similar looking communities. Unfortunately, without the label data, other direct comparisons are difficult.

\subsection{Case Studies}

\subsubsection{L\'es Miserables}

The \emph{L\'es Miserables Co-occurrence} network has $77$ nodes and $254$ edges, where a node represents a character, and an edge is weighted by the number of scenes two characters share during any chapter of Victor Hugo's novel ``L\'es Miserables''~\cite{knuth1993stanford}. The node classification comes from the primary group affiliation of the characters in the novel; the groups are named based upon our knowledge of those characters.

In \autoref{fig:teaser:before}, we show the F-R force-directed layout for this dataset.
When selecting a combination of high persistence bars and contracting the remaining ones (see \autoref{fig:teaser:barcode}), we reveal some of the key characters featured in the book, seen in \autoref{fig:teaser:both}. On two opposing sides of the layout are nodes Marius and a cluster around Valjean, two main characters, along with \'Eponine---a woman in love with Marius; Javert---the primary antagonist to Valjean; Cosette---Valjean's daughter and Marius' lover; and Toussaint---a motherly-figure assisting in raising Cosette through her childhood.

\subsubsection{Madrid Train Bombing}
\label{sec.results.train}

\begin{figure}[!b]
	\centering
    \vspace{0pt}
    {\rotatebox{90}{\subfigure[Conventional layout\label{fig.train.conventional}]{\hspace{100pt}}}}
	\includegraphics[trim=30pt 15pt 30pt 15pt, clip, width=0.85\linewidth]{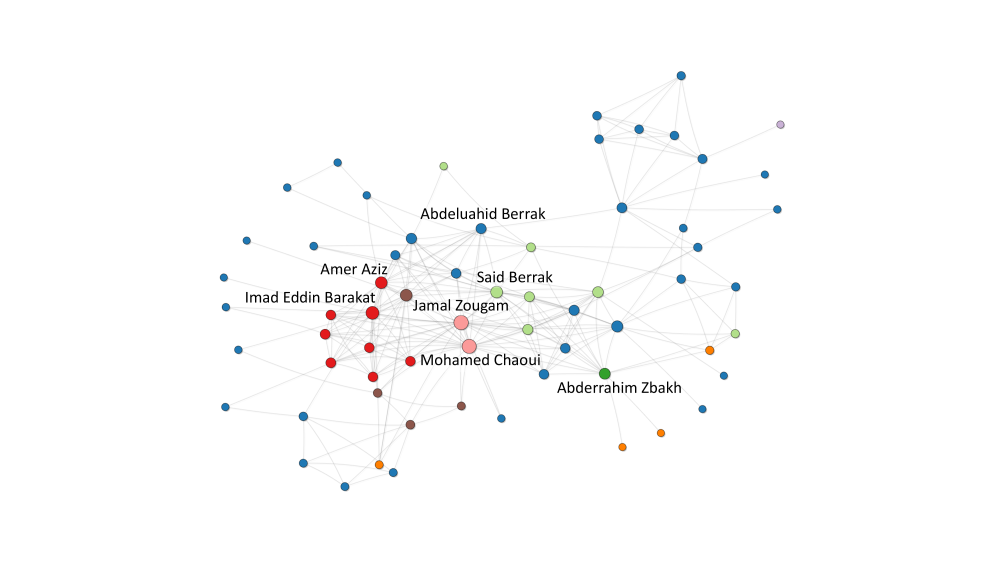}

    \vspace{0pt}
    \rotatebox{90}{\subfigure[Our approach
    \label{fig.train.final}]{\hspace{150pt}}}
	\includegraphics[trim=60pt 5pt 25pt 1pt, clip,width=0.85\linewidth]{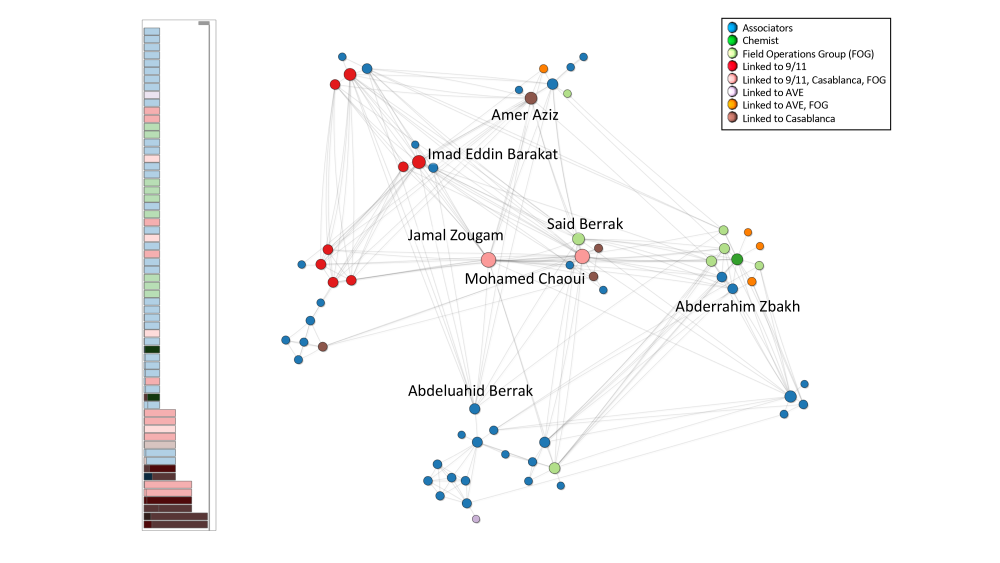}
	
	\caption{Examples from the Madrid Train Bombing dataset: (a) the conventional layout, as recreated from the original paper~\cite{rodriguez2005march}; (b) the final visualization using our layout highlighting key players in the network.}
	\label{fig.train}
\end{figure}

The Madrid Train Bombing dataset contains 70 nodes and 243 edges, where a node represents ``individuals involved in the bombing of commuter trains in Madrid on March 11, 2004''~\cite{rodriguez2005march}. Each group has been identified and colored based on whether the person was involved in previous terrorist acts and whether this person was a member of the Field Operations Group. A link is connected if two individuals were related prior to or during the bombing. Weight is calculated on an index between 1-4, where each of the following four parameters are summed per pair: trust--friendship (contact, kinship, links in the telephone center); ties to Al Qaeda and Osama Bin Laden; co-participation in training camps and/or wars; and co-participation in previous terrorist Attacks (September 11, Casablanca, etc.).

We begin by examining the layout in \autoref{fig.train.conventional}, which is drawn to replicate the original graph in Rogriguez's paper~\cite{rodriguez2005march}\footnote{We suspect some form of force-directed layout was used originally, but that information is not documented.}. This is the complete network, incorporating all players with their respective binding ties. Although Rodriguez created this graph layout to highlight the central core of the network, it is difficult to identify whom he calls the ``three most central players'' involved in the Madrid bombing.

When selecting a number of bars for contraction and repulsion, shown in \autoref{fig.train.final}, the graph closely corresponds with the analysis provided in the paper. Jamal Zougam, Mohamed Chaoui, and Said Berrak are the three most central players in the Field Operations Group network, with all three central to the graph. Below Zougam is Abdeluahid Berrak, who was suspected to be responsible for recruiting new members to the Field Operations Group. Another central player highlighted in Rodriguez's analysis is Abderrahim Zbakh, known as ``The Chemist''. In the original layout, his node is not remarkable; yet, Rodriguez described Zbakh as ``cementing the network'' by uniting unacquainted members of various terrorist groups. Other significant players include Amer Aziz and Imad Eddin Barrakat: al Qaeda, operatives closely linked as facilitators in the September 11 attacks.

\begin{figure}[!t]
	\centering
	\vspace{-5pt}
    \subfigure[2007 Co-voting\label{fig.voting.2007.c}]{{\includegraphics[height=1.35in]{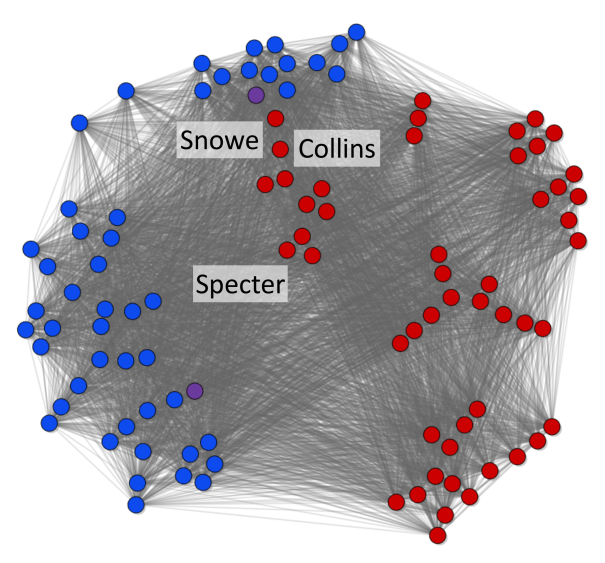}}}\hspace{10pt}
    \subfigure[2007 Anti-voting\label{fig.voting.2007.a}]{{\includegraphics[height=1.35in]{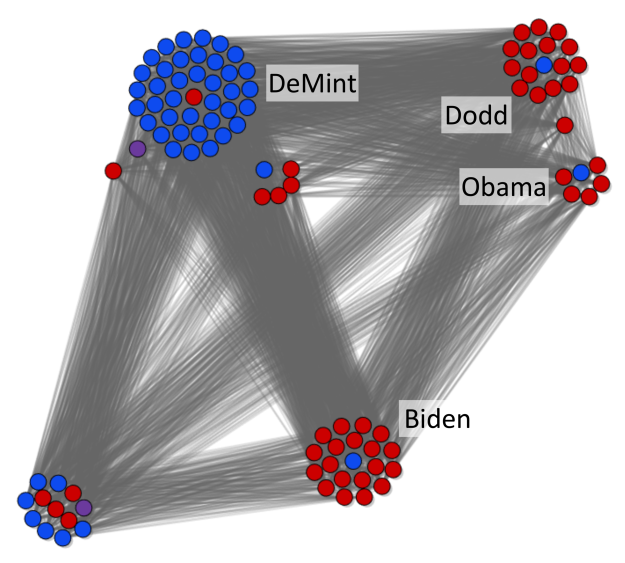}}}

    \subfigure[2008 Co-voting\label{fig.voting.2008.c}]{{\includegraphics[height=1.35in]{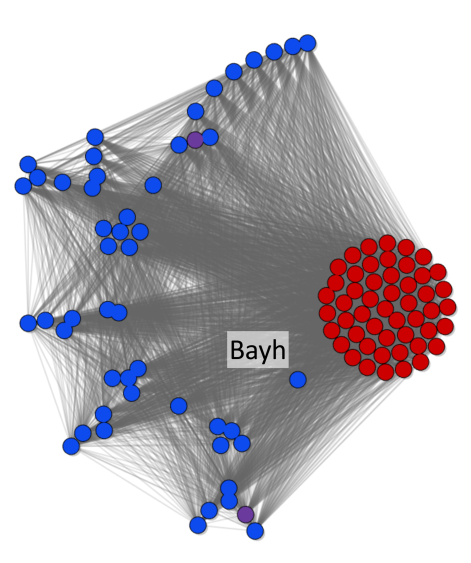}}}\hspace{10pt}
    \subfigure[2008 Anti-voting\label{fig.voting.2008.a}]{{\includegraphics[height=1.35in]{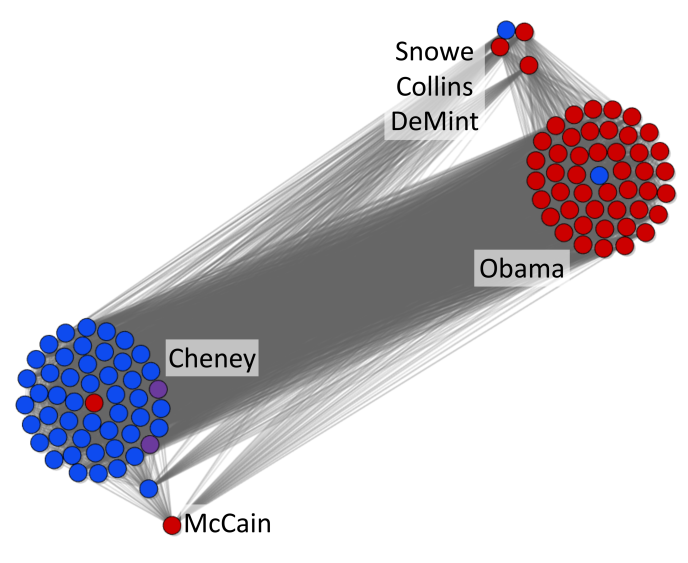}}}

    \vspace{-5pt}
	\caption{Co- and anti-voting graphs for the US Senate in 2007 and 2008. Using a mixture of contracting and repulsive forces, these graphs show the role of major political figures during these timeframes. Color are Democrats: blue; Republicans: red; and independents: purple.}
	\label{fig.voting}
\end{figure}

\subsubsection{US Senate 2007 and 2008 Co- and Anti-voting}

The \emph{US Senate 2007 and 2008} datasets are complete co- and anti-voting graphs, both with $101$ nodes (100 senators plus the Vice President) and $5,048$ edges.  This dataset is created using voting records provided by GovTrack~\cite{govtrack} with weights corresponding to how frequently two senators vote together (or against each other). 

The 2007 co-voting graph in \autoref{fig.voting.2007.c} shows the normal partisan divide---Democrats voting with Democrats and Republicans voting with Republicans. Three figures stand out among the ``centerist'' group: Snowe, Collins, and Specter. Snowe and Collins are well known for voting across partisan lines. Specter switched from the Republican to Democrat party in 2009. The 2007 anti-voting graph in \autoref{fig.voting.2007.a} highlights whom each person was most likely to vote against. The graph shows 4 clusters that isolate individuals of the opposite party: DeMint, Dodd, Biden, and Obama.

In 2008, the presidential election was in full swing, and the co-voting graph in \autoref{fig.voting.2008.c} shows that partisanship reigned. The Republicans in particular voted together. Bayh is the Democrat standout who appeared most aligned with Republicans. The 2008 anti-voting graph in \autoref{fig.voting.2008.a} highlights the politics of the election. On one side, the Republicans were running against Obama. On the Democrat side, Democrats focused on running against McCain and the Vice President, Cheney. The vote against Cheney is likely an artifact of how Congress works---the Vice President votes needs to only when there is a tie. The vote against McCain fits, since he was the Republican nominee.

\section{Discussion}

\paragraph{Transferability to Other Force-Directed Layout Algorithms.} F-R force-directed layout was used as our reference layout. Since our modification uses only additional repulsive and spring forces, by and large, other force-directed layout variants, such as sfdp~\cite{Hu2005} or FM$^3$~\cite{hachul2004drawing}, should be able to adapt our approach to their algorithms and see benefits similar to those we have demonstrated.

\paragraph{Relationship to Hierarchical Clustering.} Calculating $0$-dimensional PH features has a strong relationship to finding hierarchical clusters. However, the main differentiating factor is the treatment of features in PH. For example, PH sees low-weight edges as ``noise'', and hence collapsing them makes sense. At the same time, it sees high-weight edges as signals, and hence separating such features is meaningful.

\paragraph{(Semi-)Automatically Selecting Features.} Given the procedure defined in \autoref{sec.vis.session}, it is possible to imagine that a (semi-)automatic heuristic-based approach could be used to initialize the contraction and repulsion. We did not pursue such an approach for two reasons. First, the amount of interaction saved by such an approach would not be that significant. In our experience, exploring the contraction and repulsion options takes only a few minutes. Second and more importantly, the interaction process provides intuition about the graphs. This intuition is critical in selecting a final graph layout.

\paragraph{Disconnected Graphs.} In the case of a disconnected graph, PH calculations work without any modification by recognizing that disconnected components never merge. After that, our approach considers each connected component in the graph separately.

\paragraph{Limitations.} Our approach is not without limitations. The first problem is that force-directed layouts are, by their nature, over-constrained. Adding additional forces can exacerbate this problem. We see this in a number of examples that have negative results for contraction effectiveness (see \autoref{sec.results.layoutQ}). Second, there is the possibility for selective-engineering of the layout. Users can choose to ignore or select any PH feature to repulse, which could lead to intentionally ignoring important PH features in the final layout. Next, the worst case number of interactions needed can be quite large. To test every possible contraction and repulsion requires $2|N|$ interactions, not to mention the cost of testing combinations. Fortunately, we find the actual interaction process to be fairly quick for high-quality results (see \autoref{sec.vis.session}). Finally, our approach assumes that a unique MST exists for an input graph. If such an assumption does not hold (for example in \autoref{sec.results.train}), an arbitrary tree from the set of MSTs is selected. Selecting an \emph{optimal} MST using various constraints deserves future investigation.

\section{Conclusion}

We have presented a new approach to graph drawing that uses PH to interactively modify a force-directed layout. The approach provides a flexible interface for selecting layouts that highlight features of the graph, as defined by PH. In the future, we would like to look at the potential of using higher-dimensional PH features to control graph drawing. For example, $1$-dimensional PH features encode tunnels within metric spaces, which would be useful in constraining similar structures in the graph. However, with higher-dimensional PH features, finding the set of points that generate these features is a nontrivial problem.

\acknowledgments{We thank the reviewers for their valuable feedback. This work was supported in part by National Science Foundation grants IIS-1513616 and DBI-1661375, CRA-W Collaborative Research Experiences for Undergraduates (CREU) program, DARPA CHESS FA8750-19-C-0002, and an NVIDIA Academic Hardware Grant.}

\setstretch{1}
\bibliographystyle{abbrv-doi}

\bibliography{PH-GraphDrawing}

\end{document}